\documentclass[11pt, a4paper, onecolumn]{article}
 \usepackage{bm}
\usepackage{graphicx}
\usepackage{amssymb}
\usepackage{amsmath}
\usepackage{algorithm}
\usepackage{algpseudocode}
\usepackage{tikz}
\usepackage{booktabs}
\usepackage{tcolorbox}
\usepackage{makeidx}
\usepackage{mathrsfs}
\usepackage[left=1.6cm, right=1.6cm, top=1.6cm, bottom=2cm]{geometry}

\usepackage{makeidx}        % 先加载索引宏包
\usepackage{natbib} % 后加载 natbib
\makeindex                  % 索引命令

\usepackage{subcaption}
\usepackage{url}
\usepackage{float}

\newcommand{\R}{\mathbb{R}}

\newtheorem{theorem}{Theorem}
\newtheorem{definition}{Definition}

\newtheorem{lemma}{Lemma}
\newtheorem{remark}{Remark}
\newtheorem{proposition}{Proposition}
\newtheorem{assumption}{Assumption}

\DeclareFontFamily{OT1}{pzc}{}
\DeclareFontShape{OT1}{pzc}{m}{it}{<-> s * [1.200] pzcmi7t}{}
\DeclareMathAlphabet{\mathpzc}{OT1}{pzc}{m}{it}

\newcommand*\mcapinn[2]{\vcenter{\hbox{$\mathsurround=0pt
  \ifx\displaystyle#1\textstyle\else#1\fi\bigcap$}}}

\newcommand*\mcupinn[2]{\vcenter{\hbox{$\mathsurround=0pt
  \ifx\displaystyle#1\textstyle\else#1\fi\bigcup$}}}

\def\begequarr{\begin{eqnarray}}
\def\endequarr{\end{eqnarray}}
\def\begequarrs{\begin{eqnarray*}}
\def\endequarrs{\end{eqnarray*}}
\def\begequ{\begin{equation}}
\def\endequ{\end{equation}}
\def\begequs{\begin{equation*}}
\def\endequs{\end{equation*}}
\def\begite{\begin{itemize}}
\def\endite{\end{itemize}}

\def\begcen{\begin{center}}
\def\endcen{\end{center}}
\def\begrem{\begin{remark}\rm}
\def\endrem{\end{remark}}
\def\ba{\begin{array}}
\def\ea{\end{array}}

\def\rank{\textnormal{rank}\;}

\newcommand{\mV}{\mathrm{V}}

\newcommand{\bb}{\mathbf{b}}

\newcommand{\ub}{\mathbf{u}}
\newcommand{\vb}{\mathbf{v}}

\newcommand{\xb}{\mathbf{x}}
\newcommand{\yb}{\mathbf{y}}
\newcommand{\zb}{\mathbf{z}}

\newcommand{\etab}{\bm{\eta}}

\newcommand{\Lb}{\mathbf{L}}
\newcommand{\Ib}{\mathbf{I}}

\newcommand{\Fb}{\mathbf{F}}
\newcommand{\Eb}{\mathbf{E}}

\newcommand{\Ab}{\mathbf{A}}
\newcommand{\Bb}{\mathbf{B}}
\newcommand{\Cb}{\mathbf{C}}
\newcommand{\Db}{\mathbf{D}}

\newcommand{\eb}{\mathbf{e}}

\newcommand{\Yb}{\mathbf{Y}}

\newcommand{\gammab}{\bm{\gamma}}

\def\beeq#1{\begin{equation}{#1}\end{equation}}
\title{\huge Differential Privacy on Affine Manifolds: Geometrically Confined Privacy in Linear Dynamical Systems}

\author{%
    Zihao Ren,
    Lei Wang,
    Deming Yuan, and
    Guodong Shi\thanks{Z. Ren and G. Shi are with the Australian Centre for Robotics, The University of Sydney, NSW 2006, Australia (Email: zren0735@uni.sydney.edu.au, guodong.shi@sydney.edu.au). L. Wang is with College of Control Science and Engineering, Zhejiang University, Hangzhou, China (Email: lei.wangzju@zju.edu.cn). D. Yuan is with School of Automation, Nanjing University of Science and Technology, Nanjing, China (Email: dmyuan1012@njust.edu.cn). This work was supported by the Australian Research Council under Grant
DP190103615, Grant LP210200473, and Grant DP230101014.}
}

\date{}

\begin{document}

\maketitle

\begin{abstract}
In this paper, we present a comprehensive framework for differential privacy over affine manifolds and validate its usefulness in the contexts of differentially private cloud-based control and average consensus. We consider differential privacy mechanisms for linear queries when the input data are constrained to lie on affine manifolds, a structural property that is assumed to be available as prior knowledge to adversaries. In this setting, the definition of neighborhood adjacency must be formulated with respect to the intrinsic geometry of the manifolds. We demonstrate that such affine-manifold constraints can fundamentally alter the attainable privacy levels relative to the unconstrained case. In particular, we derive necessary and sufficient conditions under which differential privacy can be realized via structured noise injection mechanisms, wherein correlated Gaussian or Laplace noise distributions, rather than i.i.d. perturbations, are calibrated to the dataset. Based on these characterizations, we develop explicit noise calibration procedures that guarantee the tight realization of any prescribed privacy budget with a matching noise magnitude. Finally, we show that the proposed framework admits direct applications to linear dynamical systems ranging from differentially private cloud-based control to privacy-preserving average consensus, all of which naturally involve affine-manifold constraints. The established theoretical results are illustrated through numerical examples.
\end{abstract}


 \section{Introduction}\label{sec:introduction}
The rapid advances in big data and machine learning over the past decade have driven a revolution across many engineering disciplines, including manufacturing, the Internet of Things (IoT), e-commerce, healthcare, and computer vision. Powerful data representations and scalable learning algorithms increasingly unlock the value of large datasets, enabled by significant improvements in data collection, storage, and processing capabilities \cite{qiu2016survey}. As a result, information about individuals’ identities, preferences, and activities is often embedded in data collected from sensors and social media platforms. The associated privacy risks are therefore becoming a critical concern in data-driven applications \cite{zhu2020r}, particularly when datasets contain sensitive information such as political affiliations, biometric identifiers, and healthcare records \cite{wu2020value}. 

A fundamental challenge in privacy-preserving data analysis is the trade-off between protecting individual data points and maintaining analytical accuracy. Early studies \cite{denning1979tracker,denning1980fast} identified privacy threats in statistical databases, showing that adversaries can infer confidential information through carefully crafted sequential queries. It was later established that exact query responses inevitably leak information unless random noise is introduced \cite{dinur2003revealing}. Motivated by this insight, the seminal works \cite{dwork2006calibrating,dwork2014algorithmic} introduced differential privacy as a formal framework for quantifying privacy risks in randomized mechanisms. Differential privacy characterizes privacy loss through a numerical budget that bounds the similarity between the output distributions induced by adjacent inputs. A smaller privacy budget ensures that an adversary observing the output cannot confidently infer the underlying input. Differential privacy has since been successfully applied to large-scale data analysis \cite{mcsherry2009privacy} and to a wide range of fields, including deep learning \cite{yu2019differentially}, computer vision \cite{zhu2020private}, control systems \cite{yazdani2018differentially}, and distributed computation \cite{nozari2017differentially,han2016differentially,li2021privacy}.

%Differential privacy, proposed in \cite{dwork2006calibrating}, is a rigorous notion for defining and preserving data privacy. In last decades, extensive developments have been emerged in such as the mechanism design \cite{mcsherry2007mechanism} and applications to machine learning \cite{chaudhuri2011differentially} and deep learning \cite{abadi2016deep}, etc, advancing the differential privacy as a gold standard in data privacy. 

For data generated by linear dynamical systems, the system states, inputs, and control signals are inherently restricted by the temporal evolution of the system. Such restrictions  impose geometric constraints on the state space, effectively providing global knowledge about where the true system state can lie:
\begin{itemize}
\item The trajectories of the system state for a linear time-invariant system always lie in an affine subspace induced by the system equation. For cloud-based and networked control systems, eavesdroppers with access to communication packages may infer the system state, leading to an affine-manifold confined privacy problem. 
\item The agent states in a distributed consensus-seeking algorithm over a multiagent network are constrained in the affine subspace where the sum of the agent states is constant. Information about the agent initial values or states may also be leaked from the agent communications, which is also an affine-manifold confined privacy problem. 
\end{itemize}
While a substantial body of research has been developed at the interface of control theory and differential privacy (see, e.g., \cite{Ny-Pappas-TAC-2014,Hale-LQG,kawano2021modular,Lei2023,LE2025,Geir2012,NOZARI2017221,WANG2024averageconsensus}), the literature still lacks a formal and systematic framework for investigating differential privacy that explicitly accounts for the intrinsically geometrically constrained structure of data in linear systems. In this paper, we attempt to fill this gap by developing a comprehensive and systematic framework for differential privacy over affine manifolds. 

Within this framework, we explicitly account for affine-manifold constraints on the input data and examine how such structural information, assumed to be available as prior knowledge to adversaries, fundamentally alters the achievable levels of differential privacy. We analyze the impact of these constraints on privacy guarantees for linear query mechanisms and show how the underlying affine geometry can be strategically exploited to design noise-injection mechanisms that are both privacy-preserving and performance-efficient. The practical relevance and effectiveness of the proposed framework are demonstrated through its application to two representative problems: differentially private cloud-based control and differentially private average consensus. The main contributions of this paper are summarized as follows.
\begin{itemize}
    \item[(i)] We introduce a new notion of adjacency that captures the affine-manifold constraint, leading to a formulation of differential privacy over affine manifolds. For both the Gaussian and Laplace mechanisms, we establish necessary and sufficient conditions under which a prescribed differential privacy level can be achieved using structured (non–i.i.d.) noise injection.
    \item[(ii)]   For any expected privacy level, systemic approaches are developed for realizing the privacy budget sufficiently and tightly with structured noise injection. As a result, a systemic framework for differential privacy over affine manifolds with linear queries has been established.
    \item[(iii)] The obtained theories and proposed approaches are applied  to the cloud-based control problem for feedback systems, and to average consensus seeking over networks. In these applications, the affine-manifold constraint is shown to have naturally arisen  and the framework developed in this work becomes effective in differential privacy analysis and noise injection mechanism design.
\end{itemize}
These results reveal that the proposed affine-manifold constrained differential privacy and noise injection mechanisms may serve as a unified tool for data generated from linear dynamical systems.

\noindent{\bf Related Work}. Early studies investigated mechanisms based on random noise injection into measurements to protect the privacy of state trajectories, first in the context of filtering problems \cite{Ny-Pappas-TAC-2014} and later for cloud-based linear quadratic regulation \cite{Hale-LQG}. Subsequently, \cite{kawano2021modular} characterized the fundamental trade-off between control performance and the privacy of control inputs and initial states in modular tracking control, deriving explicit lower bounds on achievable differential privacy levels. The relationship between differential privacy and system observability for initial-state privacy in linear time-invariant systems was examined in \cite{Lei2023}. More recently, from a mechanism-design perspective, \cite{LE2025} investigated stochastic quantization as a privacy-enforcing mechanism, demonstrating how stochastic quantizers can be embedded within feedback loops to regulate the privacy–performance trade-off. In parallel, differentially private average consensus has been studied in \cite{Geir2012} and \cite{NOZARI2017221}, showing that perturbing inter-agent communication messages can preserve the differential privacy of agents’ initial values. Building on this line of work, \cite{WANG2024averageconsensus} designed a distributed shuffling mechanism with correlated noise to manage the privacy–accuracy trade-off in consensus outcomes. Differential privacy has also been incorporated into distributed optimization with nonlinear update rules, e.g.,  \cite{han2016differentially,Farokhi2021,XIE2023,GAO2023}.

Privacy may also be quantified in terms of the difficulty of accurately estimating sensitive parameters or signals from available observations. For example, \cite{roy2012} employed maximum likelihood estimation to assess an adversary’s inherent ability to infer system states, while \cite{mo2017} developed an optimal privacy-preserving consensus algorithm in which privacy is measured by the covariance matrix of the maximum likelihood estimate of the initial state. In \cite{FAROKHI2019275}, a privacy metric was introduced based on the inverse of the trace of the Fisher information matrix, which provides a lower bound on the variance of estimation error for unbiased estimators. In \cite{tanaka2017directed}, the causally conditioned directed information was shown to be an appropriate measure of privacy loss  from the state random variable to a random variable disclosed to the public authority for linear quadratic Gaussian control.

\noindent{\bf Paper Organization}.  The remainder of the paper is organized as follows. In Section \ref{secprob}, we introduce the system setup and define the problem of the study. In Section \ref{secresults}, we present our main results in characterizing the conditions for differential privacy with manifold dependency. Section \ref{secmech} presents algorithms designed for differential privacy over affine manifolds by generating structured (non i.i.d.) noise randomization.  Section \ref{seccontrol} moves to applications of our framework in  differentially private cloud-based control. Section \ref{secconsensus} presents  application in distributed consensus seeking, for which an algorithm with privacy and convergence guarantees is derived in view of manifold dependency.  Finally Section \ref{secconc} presents a few concluding remarks, and all proofs are provided in the Appendices.

\medskip

\noindent{\bf Notation}. We denote by $\R$ the real numbers, $\R^n$  the real space of $n$ dimension for any positive integer $n$ and $\mathbb{N}$ the set of natural numbers. For a vector $\xb\in\R^n$, we denote $x_i$ as the $i$-th entry of $\xb$, $\|\xb\|_0$, $\|\xb\|_1$ and $\|\xb\|$ as the $0$, $1$, and $2$-norm of vector $\xb$, respectively, and for any set ${\rm p}\subseteq\{1,2,\ldots,n\}$ of $l$ elements, $\xb_{{\rm p}}$  a vector of dimension $l$ with each entry being $x_{j}$ with $j\in \mathrm{p}$. 
Denote $\eb_i$ a basis vector of dimension $n$ whose entries are all zero expect the $i$-th being one. For any matrix $\Ab\in\R^{m\times n}$  and  set ${\rm p}:=\{p_1,\ldots,p_l\}\subseteq\{1,2,\ldots,n\}$ of $l$ elements, $\Ab_{{\rm p}}$  a matrix of dimension $m\times l$ with $i$-th column being the $p_i$-th column of $\Ab$, and we denote $\Eb_{\rm p}$ as a matrix of dimension $m\times l$ with each column being the basis vector $\eb_i$, $i\in \mathrm{p}$. For a symmetric and positive definite matrix $\Sigma\in\mathbb{R}^{n\times n}$, we denote $\sqrt{\Sigma}$ as a  $S\in\mathbb{R}^{n\times n}$ such that $SS^\top=\Sigma$.
{We denote by $\etab\sim\mathcal{N}(\mu,\sigma^2)^r$ if each entry in $\etab\in\mathbb{R}^r$ is i.i.d. drawn from a Gaussian distribution with mean $\mu$ and variance $\sigma^2$, and $\etab\sim\mathcal{L}(\mu,b)^r$, if each entry in $\etab\in\mathbb{R}^r$ is i.i.d. drawn from a Laplace distribution with mean $\mu$ and variance $2b^2$.
}

% The remainder of the paper is organized as follows. In Section \ref{secprob}, we introduce the system setup and define the problem of the study. In Section \ref{secresults}, we present our main results in characterizing the conditions for differential privacy with manifold dependency, and show the resulting design procedure in generating structured noise randomization for privacy protection.  Section \ref{secconsensus} presents  application in distributed consensus seeking, for which an algorithm with privacy and convergence guarantees is derived in view of manifold dependency. Section \ref{seccontrol} moves to applications of our framework in  differentially private cloud-based control. Finally Section \ref{secconc} presents a few concluding remarks.

\section{Problem Definition}\label{secprob}
In this section, starting from recalling the classical definitions of differential privacy and introducing a simple motivating example, we proceed to define the problem of interest. 

\subsection{Differential Privacy}
 Differential privacy is concerned with the privacy risk embedded in a \emph{mechanism}, denoted by   $\mathscr{M}:\mathcal{D}\rightarrow\mathcal{M}$, which is a randomized mapping from the input space $\mathcal{D}$ to the output space $\mathcal{M}$  \cite{dwork2006calibrating}.

\begin{definition}\label{def:Adj}({\em $\mu$-Adjacency})
For any two data points $\xb$ and $\xb'$ drawn from the set $\mathcal{D}\subseteq\mathbb{R}^n$, they are said to be $\mu$-adjacent with $\mu>0$, denoted by $(\xb,\xb')\in\textnormal{Adj}(\mu)$, if $\|\xb-\xb'\|_0= 1$ and  $\|\xb-\xb'\|_1\leq \mu$.
\end{definition}

\medskip

\begin{definition}\label{def:DP-1}({\em Differential Privacy})
%Let $(\mathcal{A},\mathcal{F},\mathbb{P})$ be a probability space.
A randomized mechanism $\mathscr{M}:\mathcal{D}\rightarrow\mathcal{M}$ is  $(\epsilon,\delta)$-differentially private under $\mu$-adjacency for $\epsilon\geq0,\delta\in[0,1)$,  if  for all $R\subseteq\mbox{range}(\mathscr{M})$, there holds
\begin{equation*}\label{eq:DP}
    \mathbb{P}\big(\mathscr{M}(\xb) \in R \big) \leq e^\epsilon \mathbb{P}\big(\mathscr{M}(\xb^\prime) \in R \big) +\delta
\end{equation*}
for any $(\xb,\xb')\in\textnormal{Adj}(\mu)$.
\end{definition}
 
%It is clear from Definition \ref{def:DP-1} that the standard notion of  differential privacy  is independent of how the data is generated. When the input data is dependent, such notion is not applicable as it may lead to unexpected privacy leakage.

\subsection{Mechanisms over Affine Manifolds: A Motivating Example}
We present an example to show that when the domain of secrets $\mathcal{D}$ is  an affine manifold, the inherent geometric constraint can not be ignored. 

\medskip

\noindent{\bf Example 1.} Consider a randomized mechanism
\beeq{\label{eq:ex1}
\mathscr{M}(x_1,x_2) := \begin{bmatrix}
                          x_1+\gamma_1 \\
                          x_2+\gamma_2
                        \end{bmatrix}
}
where $\gamma_1,\gamma_2$ denote the added random noises for protecting privacy of $(x_1,x_2)$, which are i.i.d. drawn from a Laplace distribution  $\mathcal{L}(0,\sigma_\gamma)$. Let $\sigma_\gamma =\mu/\epsilon$.  There are two cases.
\begin{itemize}
  \item[(i)]  [Differential Privacy over Euclidean Space] Let $\mathcal{D}=\mathbb{R}^2$.  The $\mathscr{M}$ is a standard Laplace mechanism with {$L_1$ sensitivity 1} and preserves the $(\epsilon,0)$-differential privacy of $(x_1,x_2)$ \cite{dwork2014algorithmic}.
  \item[(ii)] [Differential Privacy over Affine Manifolds] Let $$\mathcal{C}_{d}:=\{(x_1,x_2)^\top\in\mathbb{R}^2:x_1-kx_2=b\}
  $$
  with $k\in \mathbb{R}_{>0}$ and $b\in \mathbb{R}$. Thus, $x_1$ (or $x_2$) is indeed released twice at the output of $\mathscr{M}$, i.e., $x_1+\gamma_1$ and $(x_1-b)/k+\gamma_2$ (or $kx_2+b+\gamma_1$ and $x_2+\gamma_2$). Then it can be seen that $\left(\left(1+{\frac{1}{k} }\right)\epsilon,0\right)$-differential privacy of $x_1$ and $((1+k)\epsilon,0)$-differential privacy of $x_2$ are achieved tightly, by the sequential composability property of differential privacy \cite{dwork2014algorithmic}. This implies that the mechanism $\mathscr{M}$ is now $(g\epsilon,0)$-differentially private with $g:=\max\left\{1+k,1+{\frac{1}{k}}\right\}$ over $\mathcal{D}$. 
\end{itemize}
For Case (ii),   we may design probabilistically correlated zero-mean Laplacian noises as $\gamma_1=k\gamma_2$ and $\gamma_2\sim\mathcal{L}(0,\sigma_\gamma)$ with $\sigma_\gamma=\max\{1,1/k\}\mu/\epsilon$ under which $(\epsilon,0)$-differential privacy is also achieved.
 \hfill$\square$

\medskip

Example 1 is in line with the privacy frameworks of Pufferfish \cite{kifer2014pufferfish} and Blowfish \cite{he2014blowfish}, where the privacy of data with correlation/constraint is studied.  In particular, the affine manifold can be viewed as constraints about the data known by adversaries. In Blowfish \cite{he2014blowfish}, added protection against adversaries who know this constraint is shown to be possible by specialized policies. It is of interest to understand how the geometry of the affine manifold can be taken into consideration for possible added protections.  Besides being an interesting theoretical study, there are also practical motivations for exploring the affine-manifold geometry in differential privacy mechanisms.   Particularly, in dynamical systems the system state trajectories always imply a manifold dependency from the system dynamics.

\subsection{Differential Privacy over Affine Manifolds}
\label{sec-2c}
We consider the following mechanism with linear queries 
\begin{equation}\label{eq:mech-M}
\mathscr{M}(\xb) = \Fb \xb + \gammab
\end{equation}
where  $\xb \in \mathbb{R}^n$ is the input data, $\Fb$ is a matrix in  $\mathbb{R}^{m\times n}$, and $\gammab\in\mathbb{R}^m$ is a randomly  drawn noise.

\begin{definition}
Denote \begin{equation}\label{eq:Cd}
\mathcal{C}_{d}:=\big\{\xb\in\mathbb{R}^n:\ \Db\xb+\bb=0 \big\}
\end{equation}
as an affine manifold in $\mathbb{R}^n$, where $\Db\in\mathbb{R}^{q\times n}$ and $\bb\in\mathbb{R}^{q}$. 
The mechanism (\ref{eq:mech-M}) is a mechanism over $\mathcal{C}_{d}$ if   the input data $\xb$ is always taken from $\mathcal{C}_{d}$, and both $\Db$ amd $\bb$ are public.
\end{definition}

 Without loss of generality, we assume
$$\rank\left(\begin{bmatrix}\Db \cr\eb_i^\top\end{bmatrix}\right)=q+1\,,\quad \forall i\in{\rm V}.$$
This indeed indicates that $\Db$ is full-row-rank, i.e., $\rank(\Db)=q$, and none of entries in data $\xb$ is identifiable from the manifold $\mathcal{C}_{d}$. Let ${\rm V}=\{1,\ldots,n\}$. With $\rank(\Db)=q$, there exists a finite number, saying $l\in\mathbb{N}_+$, of sets ${\rm d}_j:=\{d_{j,1},\ldots,d_{j,q}\}\subseteq\mV$, $j\in\mathcal{I}:=\{1,\ldots,l\}$ such that cardinality $|{\rm d}_j|=q$,  and matrix $\Db_{{\rm d}_j}\in\mathbb{R}^{q\times q}$ is nonsingular.
We denote the set $-{\rm d}_j={\rm V}\backslash{\rm d}_j$ and can rewrite the manifold  $\mathcal{C}_{d}$ as
\[
\mathcal{C}_{d}^j:=\big\{\xb\in\mathbb{R}^n: \xb_{{\rm d}_j} = -\Db_{{\rm d}_j}^{-1}\Db_{-{\rm d}_j}\xb_{-{\rm d}_j}-\Db_{{\rm d}_j}^{-1}\bb \big\} 
\]
for $j\in\mathcal{I}$. 
We are now ready to introduce the notion of $\mu$-adjacency over  $\mathcal{C}_d$ by capturing the affine-manifold constraint in the notion of $\mu$-adjacency in Definition \ref{def:Adj}.

\begin{definition}\label{def:CAdj}({\em Manifold $\mu$-Adjacency})
We say $\xb$ and $\xb'$ to be $\mu$-adjacent over the affine manifold  $\mathcal{C}_{d}$, denoted by $(\xb,\xb')\in\textnormal{Adj}(\mu,\mathcal{C}_{d})$, if there exist $(i,j)\in{\rm V}\times\mathcal{I}$ such that $i\in{-{\rm d}_j}$,
\begin{equation}\label{eq:adj}\ba{l}
  |x_i-x_i'| \leq \mu \,\quad \\
  x_k=x_k'\,,\quad \forall k\in-{\rm d}_j\backslash\{i\} \\
  \xb_{{\rm d}_j}-\xb_{{\rm d}_j}'=-\Db_{{\rm d}_j}^{-1}\Db_{-{\rm d}_j}(\xb_{-{\rm d}_j}-\xb_{-{\rm d}_j}')\,.
\ea\end{equation}
\end{definition}

\medskip
The key intuition behind the above notion is to combine the affine-manifold constraint of the sensitive data with the original adjacency spirit in Definition \ref{def:Adj}. 
For any $(\xb,\xb')\in\textnormal{Adj}(\mu,\mathcal{C}_{d})$, we have
\begin{eqnarray}
  \xb-\xb' &=& \Eb_{-\rm d_j}(\xb_{-\rm d_j}-\xb'_{-\rm d_j})+\Eb_{\rm d_j}(\xb_{\rm d_j}-\xb'_{\rm d_j}) \nonumber \\
&\overset{(a)}{=}& [\Eb_{-\rm d_j} -\Eb_{\rm d_j}\Db_{{\rm d}_j}^{-1}\Db_{-{\rm d}_j}](\xb_{-\rm d_j}-\xb'_{-\rm d_j}) \nonumber  \\
&\overset{(b)}{=}& [\Ib_n - \Eb_{\rm d_j}\Db_{{\rm d}_j}^{-1}\Db]\Eb_{-{\rm d}_j}(\xb_{-\rm d_j}-\xb'_{-\rm d_j}) \nonumber \\
&\overset{(c)}{=}& [\Ib_n - \Eb_{\rm d_j}\Db_{{\rm d}_j}^{-1}\Db]\eb_i(x_{i}-x'_{i}) \label{eq:xxprime}
\end{eqnarray}
where  (a) is obtained using the  third equation of (\ref{eq:adj}), (b) is obtained using the equality $\Db_{-{\rm d}_j}=\Db\Eb_{-{\rm d}_j}$, and (c) is obtained using the second equation of (\ref{eq:adj}). It is noted that the negative quantity in the bracket of \eqref{eq:xxprime} characterizes the effect of the  affine-manifold constraint to the adjacency notion. When such a constraint is absent, i.e., $\Db=0$, then $\xb-\xb'=\eb_i(x_{i}-x'_{i})$, which recovers the classical notion of $\mu$-adjacency in Definition \ref{def:Adj}.
On the other hand, note that, Definition \ref{def:Adj}  requires the databases $\xb,\xb'$ to be distinct at only one entry (i.e., $\|\xb-\xb'\|_0= 1$) to capture individual's contribution to the database. However, for any two databases $\xb,\xb'\in \mathcal{C}_{d}$, it may be impossible for them to differ at only one entry (see Example 1). With ${\rm rank} (\Db)=q$, $\xb,\xb'\in \mathcal{C}_{d}$ may even differ for $q+1$ entries (see Example 2). This means individual’s contribution can no longer be looked into separately, but in groups, forming the reasoning behind Definition 4. 

To further address this issue, the following example is presented.

\medskip

\noindent {\bf Example 2.} Consider an example of publishing three data, i.e., $\Fb=\Ib_3$ which follows the manifold dependency  with $\mathcal{C}_{d}$ in \eqref{eq:Cd} with $\Db=[1,\,-2,\,0]$ and $\bb=0$. Then there are $l=2$ sets ${\rm d}_1=\{1\}$ and ${\rm d}_2=\{2\}$ such that $\Db_{{\rm d}_1}=1$ and $\Db_{{\rm d}_2}=-2$ are nonzero. It is noted that for any two data points $\xb=[x_1;x_2;x_3]$ and $\xb'=[x_1';x_2';x_3']$, if $x_1\neq x_1'$, then necessarily there holds $x_2\neq x_2'$ due to the manifold constraint. Thus, the adjacency in Definition \ref{def:Adj} is not applicable. However, according to Definition \ref{def:CAdj}, we can establish $(\xb,\xb')\in\textnormal{Adj}(\mu,\mathcal{C}_{d})$ if either of the followings is satisfied:
\begin{itemize}
  \item[(i).] $|x_1-x_1'|\leq\mu$, $x_2-x_2' = (x_1-x_1')/2$ and $x_3=x_3'$;
  \item[(ii).] $|x_2-x_2'|\leq\mu$, $x_1-x_1' = 2(x_2-x_2')$ and $x_3=x_3'$;
  \item[(iii).] $|x_3-x_3'|\leq\mu$, $x_1=x_1'$ and $x_2=x_2'$.
\end{itemize}
In view of the above cases, it can be seen that the manifold adjacency in Definition \ref{def:CAdj} successfully captures both the affine-manifold constraint  of the sensitive data and the original adjacency spirit, requiring the pair of data points to be distinct at the \emph{least} entries. Moreover, it is also noted that $(\xb,\xb')\in\textnormal{Adj}(\mu,\mathcal{C}_{d})$ neither implies nor is implied by $(\xb,\xb')\in\textnormal{Adj}(\mu)$. \hfill$\square$

We are now ready to present the notion of differential privacy over the affine manifold.
 \medskip
\begin{definition}\label{def:CDP-1}
The randomized mechanism $\mathscr{M}$ in (\ref{eq:mech-M})  is $(\epsilon,\delta)$-differentially over $\mathcal{C}_{d}$ for $\epsilon\geq0$ and $\delta\in[0,1)$,  if for all $R\subseteq\mbox{range}(\mathscr{M})$ and all $(\xb,\xb')\in\textnormal{Adj}(\mu,\mathcal{C}_{d})$,
\begin{equation} 
    \mathbb{P}\big(\mathscr{M}(\xb) \in R \big) \leq e^\epsilon \mathbb{P}\big(\mathscr{M}(\xb^\prime) \in R \big) +\delta \,.
\end{equation}
\end{definition}

With Definition \ref{def:CDP-1}, it is clear that the  sequential composition and parallel composition properties for the standard differential privacy \cite{mcsherry2009privacy} are still preserved under the affine-manifold constraint. Particularly, the resulting mechanism from a parallel composition of several $(\epsilon,\delta)$-differentially private mechanisms w.r.t. the manifold dependency maintains the $(\epsilon,\delta)$-differential privacy.  An implication of the differentially private  mechanism  $\mathscr{M}$ in (\ref{eq:mech-M}), lies in the fact that  entries in the random noise $\gammab$ may be probabilistically dependent. We introduce the following definition.
\medskip

\begin{definition}\label{def:CDP-2}
The noise $\gammab$ is probabilistically structured if there exists $\Lambda\in\mathbb{R}^{m\times r}$ with $\rank(\Lambda)=r \leq m$ such that $\gammab=\Lambda\etab$,
where the entries in $\etab\in\mathbb{R}^r$ are i.i.d.. 
\end{definition}
%\medskip

Each entry in $\etab\in\mathbb{R}^r$  may be i.i.d. drawn from a standard Gaussian distribution, i.e., $\etab\sim\mathcal{N}(0,1)^r$, rendering the mechanism $\mathscr{M}$ to be a Gaussian mechanism, or from a standard Laplace distribution, i.e., $\etab\sim\mathcal{L}(0,1)^r$, rendering the mechanism $\mathscr{M}$ to be a Laplace mechanism. The rank condition for  $\Lambda$ is without loss of generality as it guarantees a minimal value for the dimension of the random vector $\etab$ for a concise investigation of the mechanism.

In this paper, we focus on the theories and applications of the differential privacy of the mechanism $\mathscr{M}$ in (\ref{eq:mech-M}) under affine-manifold constraint and structured randomization of noises.  The objectives of this paper are outlined as follows:
\begin{itemize}
  \item[(i)] For the structured  random noise $\gammab$ under     Gaussian mechanism and Laplace mechanism, establish  conditions    under which  $(\epsilon,\delta)$-differential privacy can be achieved.
  \item[(ii).] Develop  strategic and structured noise injection methods for Gaussian and Laplace mechanisms under which a given privacy budget is tightly met in view of the affine-manifold constraint.
  \item[(iii).] Apply the established results to distributed average consensus algorithms and to cloud-based control where the affine-manifold data constraint naturally arises, and show the effectiveness of the proposed methodologies.
  \end{itemize}

%%%%%%%%%%%%%%%%%%%%%%%%%%%%%%%%%%%
\section{Differential Privacy Conditions on Manifolds} \label{secresults}
In this section, we present conditions for the mechanism $\mathscr{M}$ in (\ref{eq:mech-M}) to achieve differential privacy, under Gaussian and Laplacian mechanisms, respectively.

\subsection{Gaussian Mechanism}
\label{sec-IIIA}
First of all, we study the case with  the entries in  $\etab\in\mathbb{R}^r$  drawn  from a  standard Gaussian  distribution, i.e. $\etab\sim\mathcal{N}(0,1)^r$. We introduce  $\Lambda^{\dag}=(\Lambda^\top\Lambda)^{-1}\Lambda^\top$,  $\Psi_j:= \Ib_{n}-\Eb_{\rm d_j}(\Db_{\rm d_j})^{-1}\Db$ for $j\in\mathcal{I}$, and
\[
\Delta_i^{\mathcal{N}} = \max_{j\in\mathcal{I}} \|\Lambda^{\dag}\Fb\Psi_j\Eb_{-\rm d_j}\Eb_{-\rm d_j}^\top\eb_i\|\,,\quad i\in\mV.
\]
Denote $\Db^{\bot}\in\mathbb{R}^{n\times (n-q)}$ as a matrix such that $\Db\Db^{\bot}=0$ and $\rank\left(\begin{bmatrix} \Db^{\top} & \Db^{\bot}\end{bmatrix}\right)=n$,  and define
$$\ba{l}
\Delta_{\mathcal{N}}:= \max_{i\in\mathrm{V}}\{\Delta_i^{\mathcal{N}}\}\,,
\ea$$
and
\begin{equation}\label{eq:kappa}
    \kappa(x,y):=\Phi(\frac{y}{2}-\frac{x}{y})-e^{x}\Phi(-\frac{y}{2}-\frac{x}{y})
\end{equation} 
with $\Phi(w)=\frac{1}{\sqrt{2\pi}}\int^{w}_{-\infty} e^{-\frac{v^2}{2}}d v$. It is noted that 
\[
\frac{\partial\kappa(x,y)}{\partial y} = \frac{1}{\sqrt{2\pi}}\mbox{exp}\Big(-\frac{1}{2}(\frac{y}{2}-\frac{x}{y})^2\Big)>0\,,
\]
which implies that $\kappa(x,y)$ is a strictly increasing function with respect to $y>0$, uniformly in $x$. For convenience, given any $x$, we denote  by $\kappa^{-1}_x(\cdot)$ the inverse of the function $\kappa(x,y)$ with respect to $y$.

We present the following result.

\begin{theorem} 
\label{Theo:BDP-Gau}
  The mechanism $\mathscr{M}$ in (\ref{eq:mech-M}) with $\etab\sim\mathcal{N}(0,1)^r$ achieves $(\epsilon,\delta)$-differential privacy  over $\mathcal{C}_d$ under $\mu$-adjacency if and only if there hold
\begin{eqnarray}
  \rank(\Lambda) = \rank\left(\begin{bmatrix}\Lambda & \Fb\Db^{\bot}\end{bmatrix}\right)=r\,\label{eq:qual-bdp-G}
    ;\\
  \Delta_{\mathcal{N}}\leq \kappa_{\epsilon}^{-1}(\delta)/\mu
    \label{eq:quan-bdp-G}\, .
\end{eqnarray}
%   \begin{equation}\label{eq:qual-bdp-G}
%     \rank(\Lambda) = \rank\left(\begin{bmatrix}\Lambda & \Fb\Db^{\bot}\end{bmatrix}\right)=r\,;
%   \end{equation}
%   \begin{equation}\label{eq:quan-bdp-G}
%     \Phi\left(\frac{\mu\Delta_{\mathcal{N}}}{2}-\frac{\epsilon}{\mu\Delta_{\mathcal{N}}}\right) -e^{\epsilon}\Phi\left(-\frac{\mu\Delta_{\mathcal{N}}}{2}-\frac{\epsilon}{\mu\Delta_{\mathcal{N}}}\right) \leq  \delta\,.
%   \end{equation}
\end{theorem}

\subsection{Laplace Mechanism}

Next, we consider the Laplace mechanism with $\etab\backsim\mathcal{L}(0,1)^r$, and study the $(\epsilon,0)$-differential privacy of the mechanism $\mathscr{M}(\xb)$  w.r.t. the affine-manifold constraint $\mathcal{C}_d$. To this end,  define
$\Delta_{\mathcal{L}}:=\max_{i\in\mV} \{\Delta_i^{\mathcal{L}}\}$
with $
\Delta_i^{\mathcal{L}} = \max_{j\in\mathcal{I}} \|\Lambda^{\dag}\Fb\Psi_j\Eb_{-\rm d_j}\Eb_{-\rm d_j}^\top\eb_i\|_1$ for $i\in\mV$.

\medskip

\begin{theorem} 
\label{Theo:BDP-Lap}
The mechanism $\mathscr{M}$ in (\ref{eq:mech-M}) with $\etab\sim\mathcal{L}(0,1)^r$ achieves $(\epsilon,0)$-differential privacy over $\mathcal{C}_d$  under $\mu$-adjacency if and only if there hold
\eqref{eq:qual-bdp-G} and
  \begin{equation}\label{eq:quan-bdp-Lap}
     \Delta_{\mathcal{L}} \leq \epsilon/\mu \,.
  \end{equation}
\end{theorem}

In Theorems \ref{Theo:BDP-Gau} and \ref{Theo:BDP-Lap}, \eqref{eq:qual-bdp-G} implies the least amount of independent standard Gaussian/Laplace noises (i.e., $r\geq \rank(\Fb\Db^{\bot})$) and provides a structural property of the noise matrix $\Lambda$ for the  differential privacy; \eqref{eq:quan-bdp-G} and \eqref{eq:quan-bdp-Lap} quantify the privacy levels that can be achieved by the amount of injected Gaussian and Laplace noises, respectively.

Sensitivity analysis is a significant part when calibrating the amount of noises required to ensure a given $(\epsilon,\delta)$-differential privacy \cite{dwork2014algorithmic}.
In Theorems \ref{Theo:BDP-Gau} and \ref{Theo:BDP-Lap}, the established results on differential privacy over affine manifold $\mathcal{C}_{d}$ is also fundamentally dependent of the sensitivity analysis. Specifically, the proof of Theorems \ref{Theo:BDP-Gau} and \ref{Theo:BDP-Lap} (see Appendices A and B) lies in transforming the study of the original mechanism \eqref{eq:mech-M} over $\mathcal{C}_{d}$ to analyzing the differential privacy of mechanism $\mathscr{M}_2(\xb) = \Lambda^\dag\Fb\xb + \etab$, for which the $L_2$ sensitivity is $\Delta_2=\mu \Delta_{\mathcal{N}}$ and the $L_1$ sensitivity is  $\Delta_1=\mu\Delta_{\mathcal{L}}$, leading to the condition \eqref{eq:quan-bdp-G} for  $(\epsilon,\delta)$-differential privacy of Gaussian mechanism and the condition \eqref{eq:quan-bdp-Lap}  for $(\epsilon,0)$-differential privacy of Laplace mechanism, respectively.

We also remark that the previously established results on differential privacy of mechanism \eqref{eq:mech-M} over affine manifold $\mathcal{C}_{d}$ can  be trivially generalized to the case where  $\mathcal{C}_{d}$ is replaced by the following model
\begin{equation}\label{eq:Cd'}
\mathcal{C}_{d}':=\big\{\xb\in\mathbb{R}^n:\ \Db\xb+\bb(\Yb)=0 \big\}
\end{equation}
where $\Db$ is public and $\bb(\Yb)$ is a public mapping of $\Yb:=\mathscr{M}(\xb)$. Note that such a correlated data generation model $\mathcal{C}_{d}'$ may appear in casual systems, where $\xb$ and $\Yb$ are, respectively, a collection of sensitive data $\xb(t)$ and the observed information $\yb(t)$ by the adversary, and the sensitive data at time $t$ (i.e., $\xb(t)$) depends on the past sensitive data $\xb(t-1)$ and the observed information $\yb(t-1)$. See Sections V and VI for more explicit examples. 
In such a generalized case, after observing the value of $\Yb$, the true data distribution model $\mathcal{C}_{d}'$ is available and in fact becomes an affine-manifold model from the view point of the adversary. Besides,  as $\Yb$ is the output of the randomized mechanism $\mathscr{M}(\xb)$, no extra information can be utilized by the adversary to tell the difference between two adjacent trajectories with knowing $\bb(\Yb)$. Thus, when analyzing the differential privacy of mechanism $\mathscr{M}(\xb)$ over $\mathcal{C}_{d}'$, we can simply regard $\mathcal{C}_{d}'$ as an affine manifold and then directly apply the established results (Theorems 1 and 2) for the mechanism over affine manifold.

\section{Geometrically Structured Mechanism  Design}
\label{secmech}

In this section, we show how the conditions in Theorem 1 and Theorem 2 will inform efficient and geometrically structured randomization mechanism design.

%In this subsection, we discuss for a given privacy budget $(\epsilon,\delta)$, how we can design the structured randomization $\Lambda$ and $\etab$ so that $\mathscr{M}$ achieves $(\epsilon,\delta)$-differential privacy under $\mu$-adjacency w.r.t. the manifold correlation $\mathcal{C}_d$.

\subsection{Geometrically Structured Gaussian  Mechanism  Design}
\label{sec-GMD}
In this subsection, we discuss for a fixed Gaussian  mechanism  $\mathscr{M}$ and a given privacy budget $(\epsilon,\delta)$, how we can design the structured randomization $\Lambda$ so that $\mathscr{M}$ achieves $(\epsilon,\delta)$-differential privacy  over the affine manifold $\mathcal{C}_d$ under $\mu$-adjacency.

In Algorithm 1, we present a design approach of the Gaussian noise $\gammab:=\Lambda\etab$ by applying the two conditions \eqref{eq:qual-bdp-G} and \eqref{eq:quan-bdp-G} in Theorem \ref{Theo:BDP-Gau} such that the $\mathscr{M}$ achieves the prescribed $(\epsilon,\delta)$-differential privacy  over the affine manifold $\mathcal{C}_d$ under $\mu$-adjacency.

\begin{algorithm}[H]
{\bf Input} {Privacy levels $\epsilon\geq 0, \delta>0, \mu >0$.}

\begin{itemize}
  \item[1.] Let $\etab\sim\mathcal{N}(0,1)^r$ with $r=\rank(\Fb\Db^{\bot})$;
  \item[2.] Let $\bar\Lambda\in\mathbb{R}^{n\times r}$ be a matrix sharing the same column space with $\Fb\Db^{\bot}$;
  \item[3.] Let $\Sigma\in\mathbb{R}^{r\times r}$ be a symmetric and positive definite matrix such that
  \begin{equation}\label{eq:quan-bdp-GG}
  \Sigma \geq \left(\mu^2/{\kappa_\epsilon^{-2}(\delta)}\right)N_{ij}N_{ij}^\top \,,\quad \forall (i,j)\in\mathrm{V}\times\mathcal{J}
  \end{equation}
  with
  \begin{equation}\label{eq:bar-N_ij}
  N_{ij}:=\bar\Lambda^{\dag}\Fb\Psi_j\Eb_{-\rm d_j}\Eb_{-\rm d_j}^\top\eb_i\,.
  \end{equation}
\end{itemize}

{\bf Output} { $\gammab=\bar\Lambda \sqrt{\Sigma}\etab$.}
\caption{Structured Gaussian Noise Design Algorithm}
\label{algorithm:1}
\vspace{-0mm}
\end{algorithm}

% \begin{proposition}
%   With $\Lambda=\bar\Lambda \Sigma^{1/2}$  following Algorithm 1, the rank constraint \eqref{eq:qual-bdp-G} and the inequality \eqref{eq:quan-bdp-G} in Theorem \ref{Theo-dp} are both satisfied.
% \end{proposition}

With $\bar\Lambda$ chosen at step 2 and a nonsingular $\Sigma$ in Algorithm 1, it is clear that the design of $\Lambda=\bar\Lambda \sqrt{\Sigma}$ satisfies the rank constraint \eqref{eq:qual-bdp-G}. For the inequality \eqref{eq:quan-bdp-G}, by applying the Shur complement \eqref{eq:quan-bdp-GG} implies $N_{ij}^\top\Sigma^{-1}N_{ij}\leq \kappa_{\epsilon}^{-2}(\delta)/\mu^2$ for all $(i,j)\in\mathrm{V}\times\mathcal{J}$. This verifies \eqref{eq:quan-bdp-G}, demonstrating that the resulting Gaussian noise $\gammab=\bar\Lambda \sqrt{\Sigma}\etab$ achieves the desired $(\epsilon,\delta)$-differential privacy  over the affine manifold $\mathcal{C}_d$ under $\mu$-adjacency.

Regarding the design of $\bar\Lambda$ at the steps 1 and 2, it can be easily achieved by computing the basis vectors spanning the column space of $\Fb\Db^{\bot}$ and then assigning each column of $\bar\Lambda$ with a basis vector. For these basic matrix operations, efficient implementations are provided by most numerical linear algebra software packages. As for step 3, it involves of computing a value $\kappa_\epsilon^{-1}(\delta)$   and then designing $\Sigma$ from the linear matrix inequalities (\ref{eq:quan-bdp-GG}) for all $(i,j)\in\mathrm{V}\times\mathcal{J}$. Moreover, given $r$ and $\bar\Lambda$ at the steps 1 and 2, to minimize the covariance matrix of the added Gaussian noise $\bar\Lambda \Sigma\bar\Lambda^\top$, an optimal $\Sigma$ is derived by solving the following constrained convex optimization problem
$$\ba{ll}
\mbox{minimize}_{\Sigma>0} &\mbox{Trace}(\bar\Lambda \Sigma\bar\Lambda^\top) \\
\mbox{subject to } &\eqref{eq:quan-bdp-GG}\,.
\ea$$

%\begin{remark}\label{rem-3}
%For computing a value of $\kappa_\epsilon^{-1}(\delta)$, a numerical algorithm is developed  in \cite{balle2018improving} by transforming it into a root finding problem of a scalar monotonically increasing function, for which the Newton's method as an efficient way can be %applied to find the value to arbitrary accuracy.
% Alternatively, one may use the implication that the inequality 
% $\Phi(\frac{y}{2}-\frac{x}{y})\geq\kappa(x,y)$ by \eqref{eq:kappa} implies $\sqrt{\Phi^{-2}(\delta)+2\epsilon} + \Phi^{-1}(\delta)\geq\kappa_{\epsilon}^{-1}(\delta)$, and use $\sqrt{\Phi^{-2}(\delta)+2\epsilon} + \Phi^{-1}(\delta)$ to replace the $\kappa_{\epsilon}^{-1}(\delta)$ in \eqref{eq:quan-bdp-GG} for an analytical but less tight solution $\Sigma$.
%\end{remark}

\subsection{Structured Laplace Mechanism Design}
\label{sec-LMD}
We now turn to the Laplace mechanism $\mathscr{M}$, and propose a design approach of the random perturbation $\gammab:=\Lambda\etab$ for $(\epsilon,0)$-differential privacy by Theorem \ref{Theo:BDP-Lap}. A similar idea to Algorithm 1 is to let $\Lambda=\bar\Lambda\Sigma$ with $\bar\Lambda$ sharing column space with $\Fb\Db^{\bot}$ to satisfy the rank constraint \eqref{eq:qual-bdp-G} and the design $\Sigma$ to satisfy \eqref{eq:quan-bdp-Lap}. However, it is noted that  $\Delta_{\mathcal{L}}$ in \eqref{eq:quan-bdp-Lap} takes the form of $1$-norm, which makes it difficult to derive an explicit expression for the matrix $\Sigma$ design as in \eqref{eq:quan-bdp-GG}. In view of this, alternatively we propose to let $\Lambda=\bar\Lambda\sigma$ with a scalar $\sigma$, as shown in Algorithm 2.

\begin{algorithm}[H]
{\bf Input} {Privacy levels $\epsilon\geq 0, \delta=0, \mu >0$.}

\begin{itemize}
  \item[1.] Let $\etab\sim\mathcal{L}(0,1)^r$ with $r=\rank(\Fb\Db^{\bot})$;
  \item[2.] Let $\bar\Lambda\in\mathbb{R}^{n\times r}$ be a matrix sharing the same column space with $\Fb\Db^{\bot}$;
  \item[3.] Let $\sigma$ be such that
  \begin{equation}\label{eq:quan-bdp-LL}
    \sigma \geq \mu\bar\Delta_{\mathcal{L}}/\epsilon  \,
  \end{equation}
  with
  \begin{equation}\label{eq:bar-Delta_L}
  \bar\Delta_{\mathcal{L}}:= \max_{(i,j)\in\mathrm{V}\times\mathcal{I}}\|\bar\Lambda^{\dag}\Fb\Psi_j\Eb_{-\rm d_j}\Eb_{-\rm d_j}^\top\eb_i\|_1\,.
  \end{equation}
\end{itemize}

{\bf Output}  {$\gammab=\sigma\bar\Lambda\etab$.}
\caption{Structured Laplace Noise Design Algorithm}
\label{algorithm:2}
\vspace{-0mm}
\end{algorithm}

% \begin{proposition}
%   With $\Lambda=\bar\Lambda \Sigma^{1/2}$  following Algorithm 2, the rank constraint \eqref{eq:qual-bdp-G} and the inequality \eqref{eq:quan-bdp-Lap} in Theorem \ref{Theo-dp2} are both satisfied.
% \end{proposition}

Following Algorithm 2, it can be verified that the resulting noise matrix $\Lambda:=\sigma \bar\Lambda$ fulfills both requirements \eqref{eq:qual-bdp-G} and \eqref{eq:quan-bdp-Lap} in Theorem \ref{Theo:BDP-Lap}, achieving the desired $(\epsilon,0)$-differential privacy  over the affine manifold $\mathcal{C}_d$.
Regarding the design of $\bar\Lambda$ at the steps 1 and 2, similar to that of Algorithm 1, they mainly involve soem basic matrix operations and can be efficiently implemented by most numerical linear algebra software packages. As for step 3, it mainly involves  computing $\bar \Delta_{\mathcal{L}}$ from \eqref{eq:bar-Delta_L} which needs at most $n^2$ iterations of matrix norm computations.

\section{Differentially Private Control Systems}
\label{seccontrol}

Emerging applications in cyber-physical systems such as smart girds and intelligent transportations have inspired cloud-based control system paradigms, where a dynamical system sends its output to a cloud, and receives feedback control decisions from the cloud \cite{tanaka2017directed}. The benefit of cloud-based control is  promise in improved control accuracy and system performance since the cloud holds more information about the environment and other systems; one particular  cost   of cloud-based control is leak of  private state trajectories to adversaries eavesdropping the communication between the system and the cloud.

\subsection{Cloud-based Control Systems}

Consider the cloud-based control systems of the form
\begin{equation}\label{eq:cont-sys}
\ba{rcl}
  \xb(t+1) &=& \Ab \xb(t) + \Bb\ub(t)\,\\
  \yb(t) &=& \Cb\xb(t)
\ea\end{equation}
where the \emph{privacy-sensitive} system state $\xb(t)\in\mathbb{R}^{n_x}$, the control input $\ub(t)\in\mathbb{R}^{n_u}$, the system output/measurement $\yb(t)\in\mathbb{R}^{n_y}$.
In the cloud-enabled setup, the system transmits $\yb(t)$ to the cloud which computes a feedback control signal $\ub(t)$ and then sends it back to the system for implementation. When the communication networks between the system and the cloud are eavesdropped by adversaries, information about $\xb(t)$ might be inferred. 
%In the literature [ref],  efforts have been devoted to handle the differential privacy of states $\xb(t)$ for the dynamical system (\ref{eq:cont-sys}).

\begin{figure}[ht]
	\hspace*{0cm}
	\vspace*{0cm}
	\centering
	\includegraphics[width=8cm]{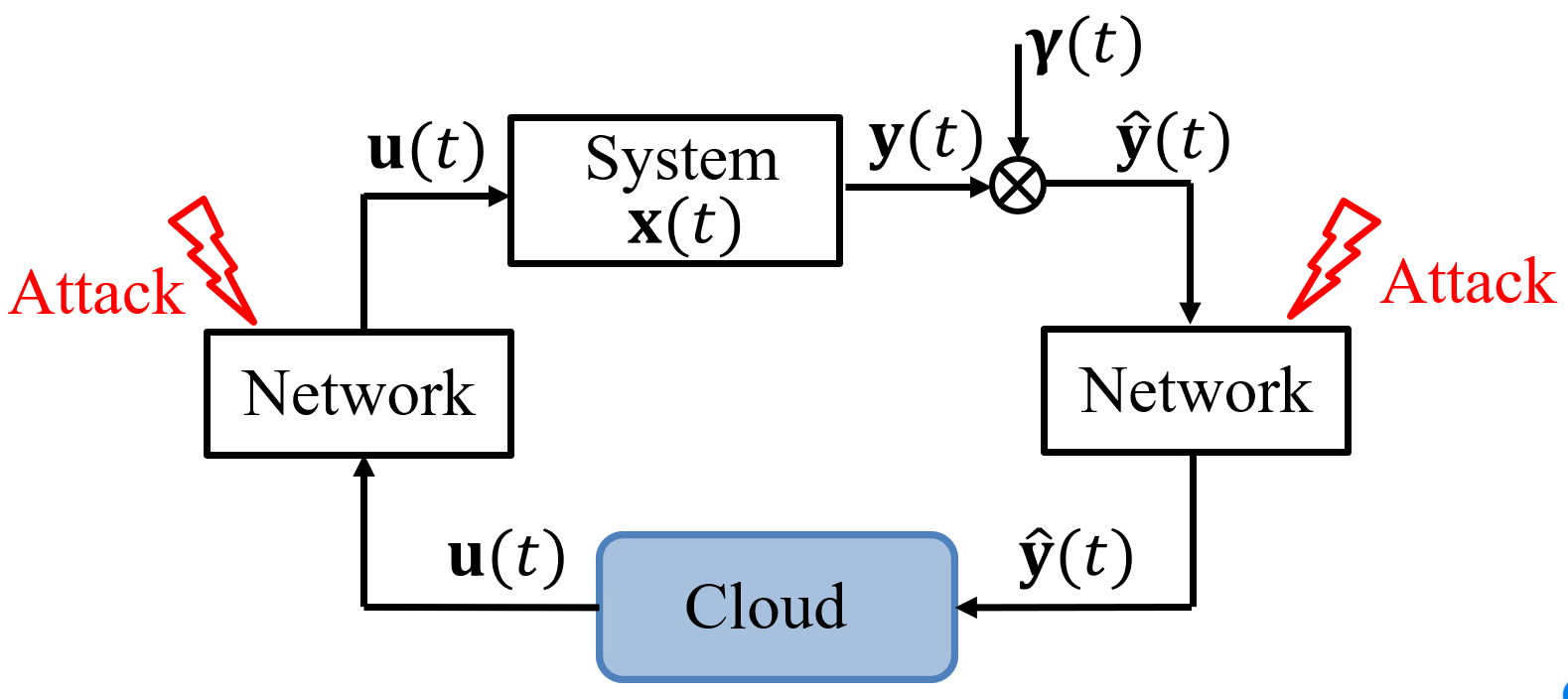}
	\caption{Differentially private cloud-based control scheme (e.g., \cite{tanaka2017directed}).}
	\label{fig:cloud_control}
\end{figure}

We propose to perturb the output with  random noises $\gamma(t)$ and submit to the cloud the perturbed output
\begin{equation}\label{eq:hat-y}
\hat\yb(t):= \Cb\xb(t) + \gamma(t)\,.
\end{equation}
See Fig. \ref{fig:cloud_control} for the considered cloud-based control scheme.
In the following, our goal is to design the random noises $\gamma(t)$ by employing the established results in Section \ref{secresults} to achieve the desired differential privacy of system states, while the readers of interest in the remainder controller design can refer to, e.g. linear–quadratic–Gaussian (LQG) control \cite{aastrom2012introduction} and neural network control \cite{ge2013stable}. As Gaussian noises are more convenient and common for tackling in these control techniques,  we choose the random noises $\gamma(t)$ as structured Gaussian noises and thus pursue the $(\epsilon,\delta)$-differential privacy with $\delta>0$.

\subsection{Differentially Private Cloud-based Control}
\label{sec-DPCbC}

Let the system running iterations $T\geq n_x$ and denote the observability matrix 
\[
\mathbf{O}_T:= [\Cb; \,\Cb\Ab ; \,\cdots; \,\Cb\Ab^{T-1} ]\,.
\]
We impose the following assumption on the system (\ref{eq:cont-sys}).

\medskip

\begin{assumption}\label{ass:full-column}
  (i). The system (\ref{eq:cont-sys}) is observable, i.e., $\rank(\mathbf{O}_T)=n_x$; (ii) The system matrix $\Ab$ in (\ref{eq:cont-sys}) is nonsingular, i.e., $\rank(\Ab)=n_x$.
\end{assumption}

%\begin{remark}
 % The observability assumption in Assumption \ref{ass:full-column} is indeed equivalent to saying that all state trajectories can be computed from trajectories of $(\yb(t),\ub(t))$ \cite{rugh1996linear}, verifying the privacy risk of $\xb(t)$. 
  %If the system (\ref{eq:cont-sys}) is non-observable, one then can always operate an observability decomposition to the system (\ref{eq:cont-sys}) \cite{rugh1996linear}, yielding a set of unobservable states $\xb_{\rm nob}(t)$ and a set of observable states $\xb_{\rm ob}(t)$, the former of which is independent of the output $\yb(t)$. Thus the privacy-preservation problem in question reduces to study the differential privacy of observable states $\xb_{\rm ob}(t)$ that obeys an observable evolution dynamics. 
  %Regarding the second part in Assumption \ref{ass:full-column}, it is made to exclude the particular case where there exist some individual entries of $\xb(t)$ that can be directly identified from the eavesdropped $\ub(t)$ by using (\ref{eq:cont-sys}). 
%\end{remark}

In the following, we apply the previous results to design the injected  random noises $\gamma(t)$ such that the differential privacy of the system state trajectory  $\big[\xb(0);\xb(1);\ldots;\xb(T-1)\big]$ is achieved under the desired privacy level $(\epsilon,\delta,\mu)$,  guaranteeing the differential privacy of state $\xb(t)$  at each time $t\in[0,T-1]$.

We define $n:=Tn_x$ and  $\xb:=\big[\xb(0);\xb(1);\ldots;\xb(T-1)\big]\in\mathbb{R}^n$, and can obtain the mechanism $\mathscr{M}$ as
\begin{equation}\label{eq:mechanism-cont}
  \mathscr{M}(\xb) = \Fb \xb + \gammab
\end{equation}
where  $\Fb = \Ib_T\otimes\Cb$ and $\gammab=\big[\gamma(0);\gamma(1);\ldots;\gamma(T-1)\big]$. Note that the private data $\xb$ is naturally and deterministically correlated by the casual system (\ref{eq:cont-sys}), i.e., subject to the  constraint
\begin{equation}\label{eq:manifold-cont}
  \mathcal{C}_d = \{\xb: \Db\xb +\bb =0\}
\end{equation}
with $\bb = (\Ib_{T-1}\otimes\Bb)\ub$, $\ub := \big[\ub(0);\ub(1);\ldots;\ub(T-2)\big]$, and
\begin{equation}\label{eq:Db}
\Db = \begin{bmatrix}
        \Ab & -\Ib_{n_x}  &  &    \\
         &   \ddots & \ddots  &\\
         &  &    \Ab & -\Ib_{n_x}
      \end{bmatrix} \in\mathbb{R}^{(n-n_x)\times n}\,.
\end{equation}
Here $\bb$ and thus the correlation model $\mathcal{C}_d$ are known by the adversary as the communication messages (i.e., $\hat\yb(t)$ and $\ub(t)$) between the system and the cloud are eavesdropped.

Moreover, there holds $\rank(\Db)=n-n_x$, and by Assumption \ref{ass:full-column},
\[
\rank\left(\begin{bmatrix} \Db \cr \eb_i^\top\end{bmatrix}\right) = n-n_x +1\,,\quad \forall i=1,\ldots,n\,.
\]

\medskip

\noindent{\bf Structured Noise Mechanism.} With the mechanism (\ref{eq:mechanism-cont}) and the affine manifold (\ref{eq:manifold-cont}), we next apply  Algorithm 1 to design the random noises $\gammab$ for a $(\epsilon,\delta)$-differentially private Gaussian mechanism $\mathscr{M}$. We denote
\begin{equation}\label{eq:D-bot}
\Db^\bot = \big[ \Ib_{n_x} ;\, \Ab  ;\, \cdots;\, \Ab^{T-1} \big]\,,
\end{equation}
satisfying $\Db\Db^\bot=0$ and $\rank\left(\begin{bmatrix} \Db^{\top} & \Db^{\bot}\end{bmatrix}\right)=n$. We define 
\[
\mathcal{O}_{jk} = \mathbf{O}_T \Ab^{1-j}\vb_k\,,\quad (j,k)\in[1,T]\times[1,n_x]\,,
\]
where  $\vb_k$ is a vector of dimension $n_x$ with entries being zero except the $k$-th being one.  Letting $\etab \sim \mathcal{N}(0,1)^{n_x}$ and $\bar\Lambda=\Fb\Db^\bot=\mathbf{O}_T$,
 we design $\Sigma$ such that
\begin{equation}\label{eq:dp-gau-ac-2}
  \Sigma \geq \left(\mu^2/{\kappa_\epsilon^{-2}(\delta)}\right)\Ab^{1-j}\vb_k(\Ab^{1-j}\vb_k)^\top \,,
  \end{equation}
for all $(j,k)\in[1,T]\times[1,n_x]$.
% \begin{equation}\label{eq:dp-gau-ac-2}
%   \Phi\Big(\frac{\mu\bar \Delta_{\mathcal{N}}}{2\sigma}-\frac{\epsilon\sigma}{\mu\bar \Delta_{\mathcal{N}}}\Big) -e^{\epsilon}\Phi\Big(-\frac{\mu\bar \Delta_{\mathcal{N}}}{2\sigma}-\frac{\epsilon\sigma}{\mu\bar \Delta_{\mathcal{N}}}\Big) \leq  \delta\,
% \end{equation}
% where $$\bar \Delta_{\mathcal{N}}=\max\limits_{(j,k)\in[1,T]\times[1,n_x]}\left\|\Ab^{1-j}\vb_k\right\|.
%$$
As a result, we design the random noise as $\gammab=\mathbf{O}_T\sqrt{\Sigma}\etab$, i.e., $
\gamma(t)=\Cb\Ab^t\sqrt{\Sigma}\etab\,, \quad \mbox{with $\etab \sim \mathcal{N}(0,1)^{n_x}$}$
for $t=0,1,\ldots,T-1$. Then the following result can be  concluded by~Theorem~\ref{Theo:BDP-Gau}.

\medskip

\begin{theorem}\label{Theo-dp2}
 Given any privacy levels $\epsilon,\delta,\mu>0$, let Assumption 1 and the random noise $\gamma(t)=\Cb\Ab^t\sqrt{\Sigma}\etab$ with $\etab\sim\mathcal{N}(0,1)^{n_x}$ and $\Sigma$ satisfying  (\ref{eq:dp-gau-ac-2}) for $t=0,1,\ldots,T-1$.
Then the system (\ref{eq:cont-sys}) with the perturbed output (\ref{eq:hat-y}) preserves the  $(\epsilon,\delta)$-differential privacy of the state trajectories under manifold $\mu$-adjacency.
\end{theorem}

%{\color{blue}
%\begin{remark}
%As shown in Theorem \ref{Theo-dp2}, if matrix $\Ab$ is an identity matrix, we then can obtain that the design of $\gammab(t)$ is independent of $T$, which implies that the proposed differentially-private cloud-based control approach is applicable even if the system running time $T$ is not fixed. However, for other cases with non-identical $\Ab$, the applicability of the proposed approach is limited to scenarios with a fixed  $T$. Despite of this limitation, we note that the proposed method is still applicable in some applications. Particularly, in some practical systems the system operator may have the knowledge of the system running time. Besides, the privacy guarantee may be expected to be against the adversary that can eavesdrop at most $T$ submitted messages. In both cases, the $T$ is fixed and  the proposed control method is applicable.
%\end{remark}
%}

\subsection{Numerical Validations}
We now provide a numerical example to illustrate the effectiveness of the privacy-preserving clould-based control approach based on the manifold differential privacy. \footnote{The codes used for producing all numerical examples in the paper are available at Github \url{https://github.com/leiwangzju/DP_affine_manifold}.}

\medskip 
\noindent{\bf System setup}. Consider the cloud-based control of autonomous vehicles with the dynamical model \cite{hoh2011enhancing,yazdani2018differentially} as  
\begin{equation}\label{eq:vehicle}\ba{rcl}
\xb(t+1) &=& A\xb(t) + B\ub(t) \\
\yb(t) &=& C \xb(t)
\ea\end{equation}
where $\xb(t)=[p(t);v(t)]$ with $p(t),v(t)$ as the \emph{private} position and the velocity, respectively, and the system matrices
\[
A=\begin{bmatrix} 1 & T_s \cr 0 & 1 \end{bmatrix}\,,\quad B=\begin{bmatrix} T_s^2/2 \cr T_s\end{bmatrix}\,,\quad C=\begin{bmatrix} 1 & 0\end{bmatrix}
\]
with the sampling period $T_s=0.1$. In the cloud-based setup, for privacy concern the vehicle sends perturbed output $\hat\yb(t)=\yb(t)+\gamma(t)$ to the cloud, which then delivers to the vehicle for implementation a control signal of the form 
\footnote{Here for simplicity we choose a classical output-feedback controller. However, it is noted that other methods such as LQG control \cite{aastrom2012introduction} and neural network control \cite{ge2013stable} are also applicable with no influence on the design of noises $\gammab(t)$ for privacy preservation.}
\begin{equation}\label{eq:ofc}\ba{rcl}
  \ub(t) &=& -K_P^\top (\hat\xb(t)-\xb_r(t))\\
  \hat \xb(t+1) &=& A \hat\xb(t) + B\ub(t) + L(\hat\yb(t)-C\hat\xb(t))
\ea\end{equation}
 with $K_P=[3.4240;4.3095]$, $L=[0.8266;0.6973]$, and the reference trajectory $\xb_r(t):=(p_r(t);v_r(t))= \big(\tanh(t);1-|\tanh(t-9)|\big)$.

\medskip 
\noindent{\bf Experiments}.
We note that the $v(t)$-dynamics in (\ref{eq:vehicle}) is dependent of $v(t),\ub(t)$ but independent of the private state $p(t)$. In other words, the privacy-sensitive positions $p(t)$ are affine-manifold-constrained on the $p(t)$-dynamics. In view of this, we apply the previous Gaussian mechanism design in Subsection \ref{sec-DPCbC} with $\Ab=1$ and $\Cb=1$, and by Algorithm 1, take  $\gamma(t)=\sqrt{\Sigma}\etab$ with $\etab\sim\mathcal{N}(0,1)$ and $\Sigma$ satisfying 
\[
  \Sigma\geq \mu^2/\kappa_\epsilon^{-2}(\delta)\,.
\]
In simulations, we let the running time $T=100$ and the desired privacy levels $(\epsilon,\delta,\mu)=(1,10^{-2},1)$, and then design $\gamma(t)=\sqrt{\Sigma}\eta$ with the standard Gaussian noise $\etab\sim\mathcal{N}(0,1)$ and $\sqrt{\Sigma}=2.5244$.

 \begin{figure}[ht]
      \centering
      \begin{subfigure}[b]{0.47\textwidth}
          \centering
          \includegraphics[width=0.98\textwidth]{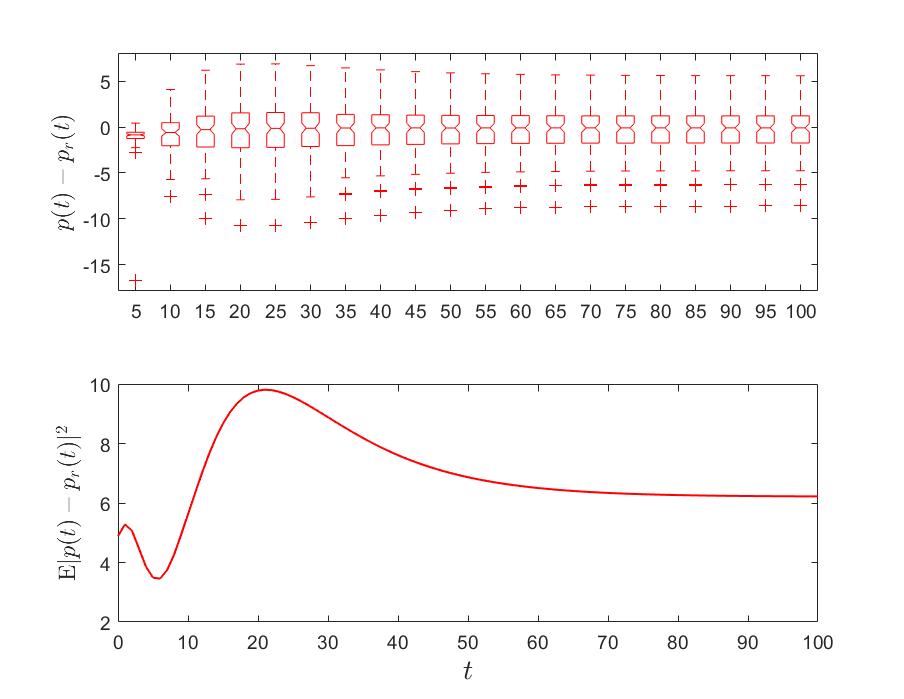}
          %\caption{}
	      \label{fig:tracking_error}
      \end{subfigure}
      \begin{subfigure}[b]{0.47\textwidth}
          \centering
          \includegraphics[width=0.98\textwidth]{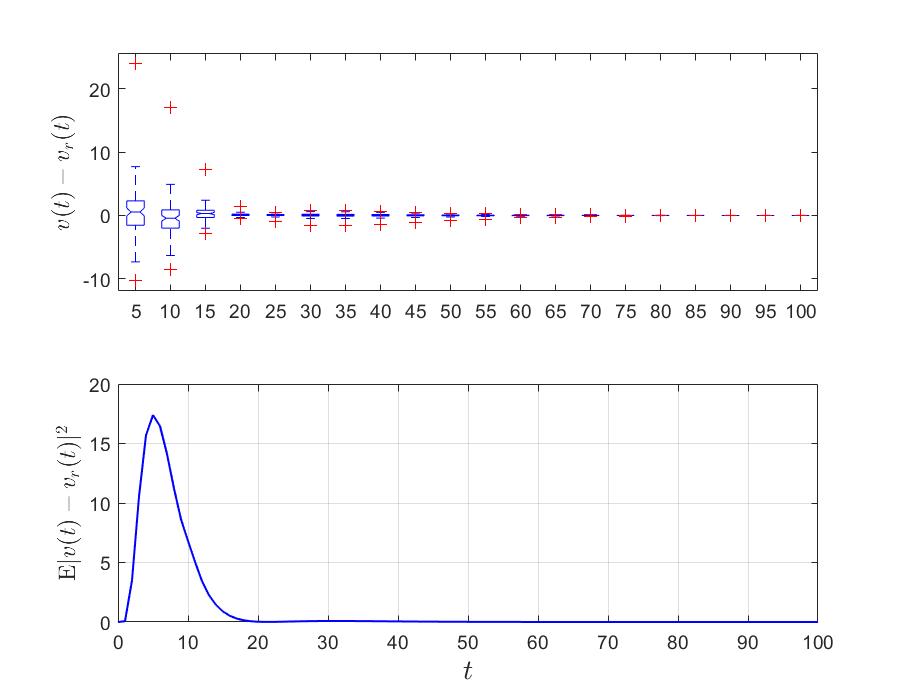}
          %\caption{}
          \label{fig:tracking_error_mean}
      \end{subfigure}
        \caption{Tracking errors $p(t)-p_r(t)$  and $v(t)-v_r(t)$. }
        \label{fig:cloud}
 \end{figure}
 
% {\color{blue}
 %\begin{remark}
 %In \cite{xiao2015protecting}, the temporal correlation of locations is modelled by the Markov chain, and the notion of $\delta$-location set is introduced to define a probabilistic version of data adjacency for differential privacy analysis. In contrast with \cite{xiao2015protecting}, we note that the temporally correlated locations in the above example are generated by the motion dynamics of vehicles, which is a casual system with the control input and cannot be modelled by the Markov chain. This demonstrate the main difference between the temporal correlation modelled by the Markov chain in  \cite{xiao2015protecting} and our affine-manifold correlation model due to the casual motion dynamics of vehicles. 
 %\end{remark}
 %}  
 
  \begin{figure}[ht]
      \centering
      \begin{subfigure}[b]{0.45\textwidth}
          \centering
          \includegraphics[width=0.99\textwidth]{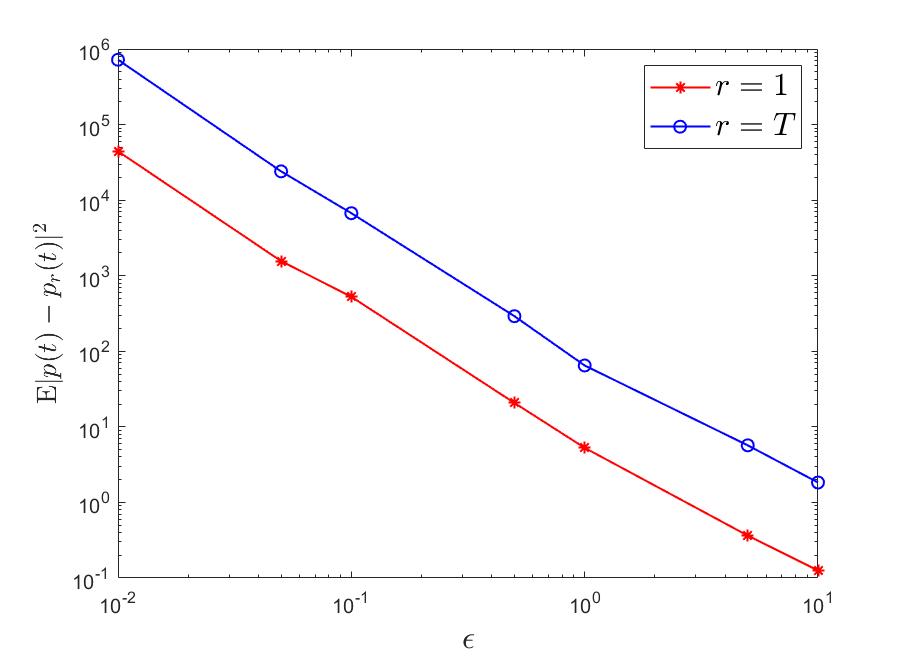}
          %\caption{}
	      %\label{fig:tracking_error}
      \end{subfigure}
      \begin{subfigure}[b]{0.45\textwidth}
          \centering
          \includegraphics[width=0.99\textwidth]{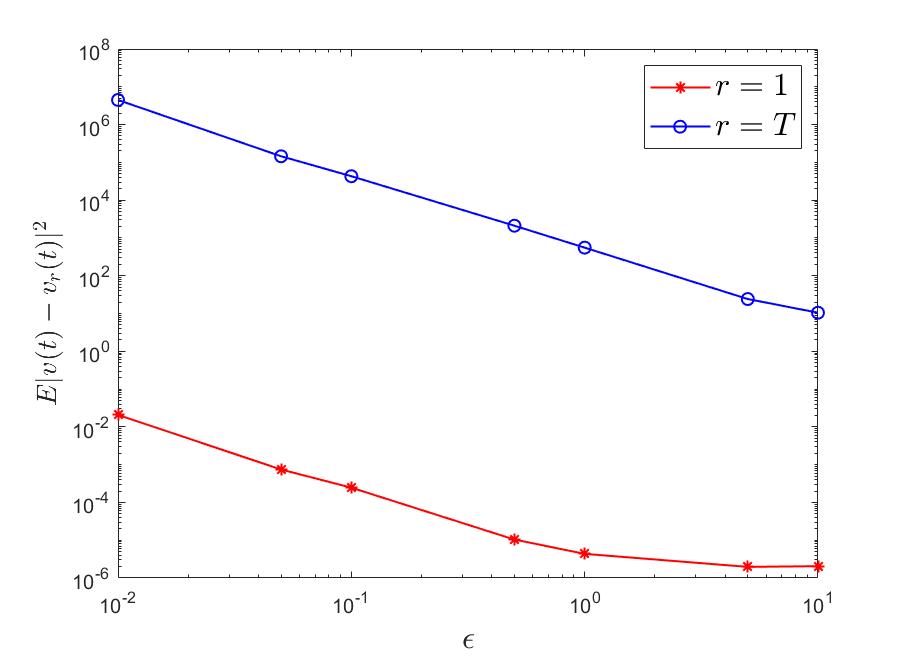}
          %\caption{}
          %\label{fig:tracking_error_mean}
      \end{subfigure}
        \caption{Mean-square tracking errors $p(t)-p_r(t)$  and $v(t)-v_r(t)$   under different privacy requirements $\epsilon$ and structured noises ($r=1$ for the structured noises following Algorithm 1 and $r=T$ for i.i.d. noises).}
        \label{fig:cloud-2}
 \end{figure}
 
\medskip  
\noindent{\bf Results}. The resulting tracking errors are presented in Fig. \ref{fig:cloud}, where the tracking velocity error is asymptotically vanishing, while the tracking position error is not vanishing in the mean-square sense due to presence of noises for privacy preservation. We also show the relationship between the tracking performance and the privacy requirements. The simulation results are presented in Fig. \ref{fig:cloud-2}. It is clear that a higher privacy level (i.e., a smaller $\epsilon$) leads to larger tracking errors of both position and velocity in the mean square senses, demonstrating the trade-off between the data utility and the differential privacy guarantee. Besides, in Fig. \ref{fig:cloud-2} we also compare the tracking performance by calibrating the noises via the proposed structured noise injection approach ($r=1$) to the common approach of using i.i.d. noises ($r=T$). 
It can be seen that  the calibrated noises following Algorithm 1 shows better tracking performances under the same privacy requirement, implying that the proposed structured noise injection approach may improve the data utility. %{\color{blue}The reason behind such an improvement comes from adding less amount of noises by improving the sensitivity of the mechanism.  As addressed in Remark 5 and also in the proof of Theorems 1 and 2, the analysis of differential privacy of mechanism over affine manifold can be transformed to analyzing the differential privacy of mechanism $\mathscr{M}_2(\xb) = \Lambda^\dag\Fb\xb + \etab$. By calibrating the noises via the proposed structured noise injection approach, we can obtain $\mathscr{M}_2(\xb) = \frac{1}{T\sigma}\mathbf{1}_T\xb + \etab$ whose $L_2$ sensitivity is $\Delta_2=1/\sigma$, while for the approach of adding i.i.d. noises, we have $\mathscr{M}_2(\xb) = \frac{1}{\sigma}\xb + \etab$ whose $L_2$ sensitivity is $\Delta_2=\sqrt{T}/\sigma$. This implies that a smaller $\sigma$ (i.e., less amount of noises) is needed using structured noises to preserve  the same differential privacy than using i.i.d. noises. }

 \section{Differentially Private Average Consensus}\label{secconsensus}

In this section, we apply the previously established results to develop an average consensus algorithm that preserves the differential privacy of node states, against the eavesdroppers having access to all communication messages. Consensus algorithms have become a standard information aggregation routine in distributed algorithms for convex optimization and machine learning \cite{vanhaesebrouck2017decentralized,nedic2010constrained,scaman2019optimal}; privacy-preserving consensus algorithms may be used to realize  distributed optimization procedure with differential privacy guarantees \cite{nozari2017differentially,han2016differentially,li2021privacy}.

\subsection{The Algorithm}

We consider a multi-agent network with $n$ agents indexed in the set $\mathrm{V}=\{1,\dots,n\}$, and suppose that there is an undirected and connected  communication network among these agents, denoted as $\mathrm{G}=(\mathrm{V},\mathrm{E})$ with $\mathrm{E}\subseteq\mathrm{V}\times\mathrm{V}$ representing the set of edges.  The proposed differentially private average consensus algorithm takes the form
\begin{equation}\label{eq:AveCon}\ba{rcl}
  x_i(t+1) &=& x_i(t) + u_i(t)\,\\
  u_i(t) &=& \sum_{j\in\mathrm{V}}w_{ij}\big(y_j(t)- y_i(t)\big)\,\\
  y_i(t) &=& x_i(t) + \gamma_i(t)
  \ea
\end{equation}
where $x_i(t)$ and $y_i(t)$ denote the privacy-sensitive state and the perturbed communication message of node $i$, respectively at time $t$, $\gamma_i(t)$ denotes the random noise added in the communication message for privacy preservation at $t$ and will be designed later. Here $w_{ij}$ denotes the weight of the edge $(i,j)$, satisfying $w_{ij}=w_{ji}>0$ if $(i,j)\in\mathrm{E}$, $w_{ji}=0$ if $(i,j)\notin\mathrm{E}$ and $\sum_{j\in\mathrm{V}}w_{ij} <1$ for all $i\in\mathrm{V}$.
In order to preserve the distributed nature of (\ref{eq:AveCon}), we let each node $i$ independently generate its local perturbation sequence $\{\gamma_i(t)\}$, i.e., for any two nodes $i,j\in\mathrm{V}$ their perturbation sequences $\{\gamma_i(t)\}$ and $\{\gamma_j(t)\}$ are mutually independent.

Let $T$ be the running steps of the algorithm (\ref{eq:AveCon}), which is allowed to be arbitrarily large and not necessarily fixed in the sequel, and denote the state trajectory of node $i\in\mV$ as $\xb_i:=[x_i(0);x_i(1);x_i(T-1)]$. It is clear that each node state trajectory $\xb_i$ is naturally correlated by the casual system \eqref{eq:AveCon}, and the mechanism for the differential privacy of node state trajectories $\xb:=[\xb_1;\ldots;\xb_n]$ can be described by
\begin{equation}\label{eq:M_Ave}
  \mathscr{M}(\xb) = \begin{bmatrix} \mathscr{M}_1(\xb_1)\cr \vdots \cr \mathscr{M}_n(\xb_n)\end{bmatrix}\,
\end{equation}
with \begin{equation}\label{eq:M_i}
\mathscr{M}_i(\xb_i) = \Fb_i \xb_i + \gammab_i\,,\quad i\in\mathrm{V}
\end{equation}
where $\Fb_i = \Ib_{T}$ and $\gammab_i=[\gamma_i(0);\ldots;\gamma_i(T-1)]\in\mathbb{R}^{T}$.

For each $i\in\mathrm{V}$ the sensitive data $\xb_i$ is correlated over the  manifold
\begin{equation}\label{eq:Cd_i}
\mathcal{C}_{d,i}:=\{\xb_i: \Db_i\xb_i+\bb_i=0\}\,
\end{equation}
where
\[\ba{l}
\Db_i = \begin{bmatrix}
        1 & -1 &  &    &\\
         & 1 & -1 &    &\\
         &  & \ddots & \ddots  &\\
         &  &  &  1 & -1
      \end{bmatrix} \in\mathbb{R}^{(T-1)\times T}\,,\\
\bb_i = [u_i(0);\ldots;u_i(T-2)]\,
\ea\]
with $\Db_i$ full-row-rank and $\bb_i$ is known by the adversary.

Bearing in mind the mechanism (\ref{eq:M_Ave})-(\ref{eq:M_i}) and the dependency manifold (\ref{eq:Cd_i}), it is worth noting that the mechanism $\mathscr{M}$ achieves the $(\epsilon,\delta)$-differential privacy if so does each $\mathscr{M}_i$ for $i\in\mV$, due to the parallel composability property of the differential privacy \cite{mcsherry2009privacy}.
Thus, we next design the random noise $\gammab_i$ for each $i\in\mV$ using either Algorithm 1 for Gaussian mechanism or Algorithm 2 for Laplace mechanism, such that the resulting mechanism $\mathscr{M}_i$ achieves the desired $(\epsilon,\delta)$-differential privacy under manifold $\mu$-adjacency.

\medskip\noindent
{\bf Gaussian Mechanism.} We first apply Algorithm 1 to design the random noise $\gammab_i$ for a Gaussian mechanism $\mathscr{M}_i$ to achieve the $(\epsilon,\delta)$-differential privacy with $\delta>0$. Let $\Db_i^{\bot}=\mathbf{1}_{T}$, which satisfies $\Db_i\Db_i^{\bot}=0$ and $\rank\left(\begin{bmatrix} \Db_i^{\top} & \Db_i^{\bot}\end{bmatrix}\right)=T$. Then we observe that $\rank(\Fb_i\Db_i^{\bot})=1$ with $\Fb_i\Db_i^{\bot}=\mathbf{1}_{T}$, and by following the steps 1 and 2 in Algorithm 1, thus let
\begin{equation}\label{eq:etab_Ave}
  \etab_i \sim \mathcal{N}(0,1)\,,\quad \bar\Lambda_i=\mathbf{1}_{T}\,.
\end{equation}
Then we follow the step 3 in Algorithm 1 and proceed to design $\Sigma_i$ such that
\begin{equation}\label{eq:dp-gau-ac}
          \Sigma_i:=\sigma_i^2\geq \mu^2/\kappa_{\epsilon}^{-2}(\delta)\,,\qquad i\in\mathrm{V}.
\end{equation}

With fixing $\etab_i,\bar\Lambda_i$ and $\Sigma_i$ in (\ref{eq:etab_Ave}) and (\ref{eq:dp-gau-ac}), we finally can complete the design of $\gammab_i$ for the Gaussian mechanism $\mathscr{M}_i$ as $\gammab_i = \mathbf{1}_{T} \sigma_i\etab_i$, i.e., $\gamma_i(t)=\sigma_i\etab_i$ for $i\in\mV$ and $t\geq 0$.

\medskip\noindent
{\bf Laplace Mechanism.}
We then apply Algorithm 2 to design the random noise $\gammab_i$ for a Laplace mechanism $\mathscr{M}_i$ such that the $(\epsilon,0)$-differential privacy can be achieved.
The application procedure is similar to the above design for the Gaussian mechanism, and is thus omitted for simplicity. As a result, the random noise $\gammab_i$ for the Laplace mechanism $\mathscr{M}_i$ is given by $\gammab_i = \sigma_i\mathbf{1}_{T}\etab_i$ with $\etab_i \sim \mathcal{L}(0,1)$ and
\begin{equation}\label{eq:dp-lab-ac}
   \sigma_i \geq  \mu/\epsilon\,, \qquad i\in\mathrm{V}.
\end{equation}
Namely, $\gamma_i(t)=\sigma_i\etab_i$ for $i\in\mV$ and $t\geq0$.

%Again, by $\Fb_i\Db_i^{\bot}=k\mathbf{1}_{T}$ for $k\neq 0$, according to steps 1 and 2 of Algorithm 2 we let
%\begin{equation}\label{eq:etab_lab-Ave}
%  \etab_i \sim \mathcal{L}(0,1)\,,\quad \bar\Lambda_i=\mathbf{1}_{T}\,.
%\end{equation}
%Then we follow the step 3 in Algorithm 2 and proceed to design $\sigma_i$ such that
%\begin{equation}\label{eq:dp-lab-ac-1}
%  \sigma \geq \mu\bar \Delta_{\mathcal{L},i}/\epsilon\,
%\end{equation}
%where
%  \[
%  \bar \Delta_{\mathcal{L},i}:=\max_{(j,t)\in\mathcal{T}\times\mathcal{T}} \|\bar\Lambda^{\dag}\Fb_i\Psi_{ij}\Eb_{-{\rm d}_{i,j}}\Eb_{-{\rm d}_{i,j}}^\top\eb_t\|_1\,.
%  \]
%Again, by recalling Lemma 2 in  \ref{app:proof_Theo3}, it can be seen that  $\bar \Delta_{\mathcal{L},i}=1$ for $i\in\mV$, which simplifies the design requirement (\ref{eq:dp-lab-ac-1}) of $\sigma_i$ as
%\begin{equation}\label{eq:dp-lab-ac}
%   \sigma_i \geq  \mu/\epsilon\,\qquad i\in\mathrm{V}.
%\end{equation}
%
%With fixing $\etab_i,\bar\Lambda_i$ and $\sigma_i$ in (\ref{eq:etab_lab-Ave}) and (\ref{eq:dp-lab-ac}), we finally can complete the design of $\gammab_i$ for the Laplace mechanism $\mathscr{M}_i$ as $\gammab_i = \sigma_i\mathbf{1}_{T}\etab_i$, i.e., $\gamma_i(t)=\sigma_i\etab_i$ for $i\in\mV$ and $t=0,1,\ldots,T-1$.

\subsection{Privacy and Convergence}

%In this subsection, the differential privacy and the convergence property of the the proposed average consensus algorithm \eqref{eq:AveCon} are studied.

The following result demonstrates that the noise design method given in the previous subsection enables the algorithm \eqref{eq:AveCon} to achieve the desired $(\epsilon,\delta)$-differential privacy, whose proof is a straightforward verification of the conditions in Theorems \ref{Theo:BDP-Gau} and \ref{Theo:BDP-Lap} and is thus omitted for simplicity.
\begin{theorem}\label{Theo-dp}
Let either of the following hold.
\begin{itemize}
  \item[(i)] \emph{(Gaussian Mechanism)} Let the privacy levels $\epsilon,\delta,\mu>0$, and the random noise $\gamma_i(t)=\sigma_i\etab_i$ with $\etab_i\sim\mathcal{N}(0,1)$ and $\sigma_i$ satisfying  (\ref{eq:dp-gau-ac}), for $i\in\mV$ and $t\geq 0$.
  \item[(ii)] \emph{(Laplace Mechanism)} Let the privacy levels $\epsilon,\mu>0$ and $\delta=0$, and the random noise $\gamma_i(t)=\sigma_i\etab_i$ with $\etab_i\sim\mathcal{L}(0,1)$ and $\sigma_i$ satisfying (\ref{eq:dp-lab-ac}), for $i\in\mV$ and $t\geq 0$.
\end{itemize}
Then the algorithm \eqref{eq:AveCon} preserves  $(\epsilon,\delta)$-differential privacy of all node state trajectories under manifold $\mu$-adjacency.
%  Consider the privacy-preserving average consensus algorithm \eqref{eq:AveCon} with the message perturbation noises $\gammab_i$ following Algorithm 1. Then the algorithm \eqref{eq:AveCon} preserves the desired $(\epsilon,\delta)$-differential privacy of all node state trajectories under manifold $\mu$-adjacency.
\end{theorem}

Let $\Lb$ be the Laplacian matrix of the communication network $\mathrm{G}$, satisfying $[\Lb]_{ii}=-\sum_{j\in\mathrm{V}}w_{ij}$ and $[\Lb]_{ij}=w_{ij}$ for $i\neq j$, $i,j\in\mathrm{V}$. Then we arrange the eigenvalues of $\Lb$ in an increasing order as $\lambda_1<\lambda_2\leq\ldots\leq\lambda_n$. There hold $\lambda_1=0$ and $|\lambda_i|<2$ for all $i=2,\ldots,n$ \cite{mesbahi2010graph}. Denote $\alpha:=\max\{|1-\lambda_2|,|1-\lambda_n|\}$, $\xb(t)=[x_1(t);\ldots;x_n(t)]$, and the average of initial states $\bar x_0 = \mathbf{1}_n^\top\xb(0)/n$. We also  present the convergence and computation accuracy results for the average consensus algorithm \eqref{eq:AveCon}.

\begin{proposition}\label{Theo-ConAcu}
  Consider the privacy-preserving average consensus algorithm \eqref{eq:AveCon} and let $\gamma_i(t)=\sigma_i\etab_i$ with  $\etab_i\sim\mathcal{N}(0,1)$ or $\etab_i\sim\mathcal{L}(0,1)$  for  $t\geq 0$ and $i\in\mathrm{V}$. Then 
  \begin{itemize}
    \item[(i)]  $\|\xb(t) - \mathbf{1}_n\bar x_0\| \leq \alpha^t\|\xb(0)\| + (1+\alpha^t)\big(\sum_{i\in\mathrm{V}}\sigma_i^2\etab_i^2\big)^{1/2}$.
    \item[(ii)]  $\lim\limits_{t\rightarrow\infty}\mathbb{E}\big[\xb(t) -  \mathbf{1}_n\bar x_0 \big]=0$.
    \item[(iii)] there holds
    \[
    \lim\limits_{t\rightarrow\infty}\mathbb{E}\|\xb(t) -  \mathbf{1}_n\bar x_0   \|^2 \leq \left\{\begin{array}{l}\sum\limits_{i\in\mathrm{V}}\sigma_i^2\,,\,\mbox{for }\etab_i\sim\mathcal{N}(0,1) \\ \sum\limits_{i\in\mathrm{V}}2\sigma_i^2\,, \,\mbox{for }\etab_i\sim\mathcal{L}(0,1)\end{array}\,.\right.
    \]
  \end{itemize}
\end{proposition}

From Theorem \ref{Theo-dp} and Proposition \ref{Theo-ConAcu}, it is clear that there is a trade-off between the differential privacy level and mean-square computation accuracy. We note that as in other  average consensus algorithms \cite{nozari2017differentially} with differential privacy guarantees of initial states, such a trade-off cannot be removed.
 However, given any mean-square error accuracy requirement $\zeta$, the optimal trade-off can be derived in terms of maximizing the privacy level, i.e., minimizing $\epsilon$ under fixed $\delta,\mu>0$ by combining Theorem 4 and Proposition 1. Specifically, the following statements can be verified. 
\begin{itemize}
    \item[(i)] For Gaussian mechanism, the optimal privacy level $\epsilon_{op}$ is given by 
\[
\epsilon_{op}=\inf\{\epsilon\geq0: \kappa(\epsilon,\mu\sqrt{{n}/{\zeta}}) \leq \delta\}\,.
\]
\item[(ii)] For Laplace mechanism, the optimal privacy level $\epsilon_{op}$ is given by
\[
\epsilon_{op} =  \mu\sqrt{{2n}/{\zeta}}\,.
\]
\end{itemize}
 
\subsection{Numerical Validations}
We now provide a numerical example validating the effectiveness of the proposed privacy-preserving consensus algorithm. 
%The code used for producing this numerical example is available at Github ???.

\begin{figure}[ht]
	\centering
	\includegraphics[width=5cm]{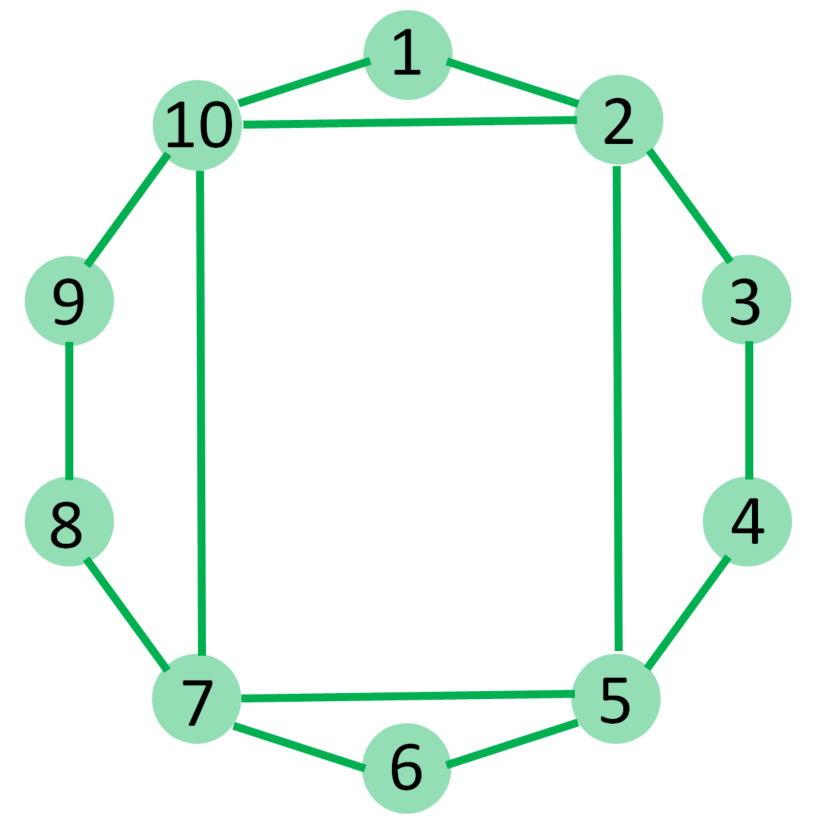}
	\caption{Communication graph $\rm G$.}
	\label{fig:network}
\end{figure}

\medskip 
\noindent{\bf System Setup}. We consider a network of 10 nodes with a communication graph given in Fig. \ref{fig:network}, in which each edge is assigned with the same weight $1/4$. Each node $i$ holds an initial state $x_i(0)$, $i=1,\ldots,10$, described by
$
\xb(0) =[10;100;20;-30;-20;-60;70;0;80;-20]\,.
$
It is clear that the average of initial values is $x^{\star}=10$. We apply the proposed differentially private average consensus algorithm (\ref{eq:AveCon}) to cooperatively compute the average while preserving the differential privacy of the node state trajectories.

\begin{figure}[ht]
	\hspace*{0cm}
	\vspace*{0cm}
	\centering
	\includegraphics[width=8cm]{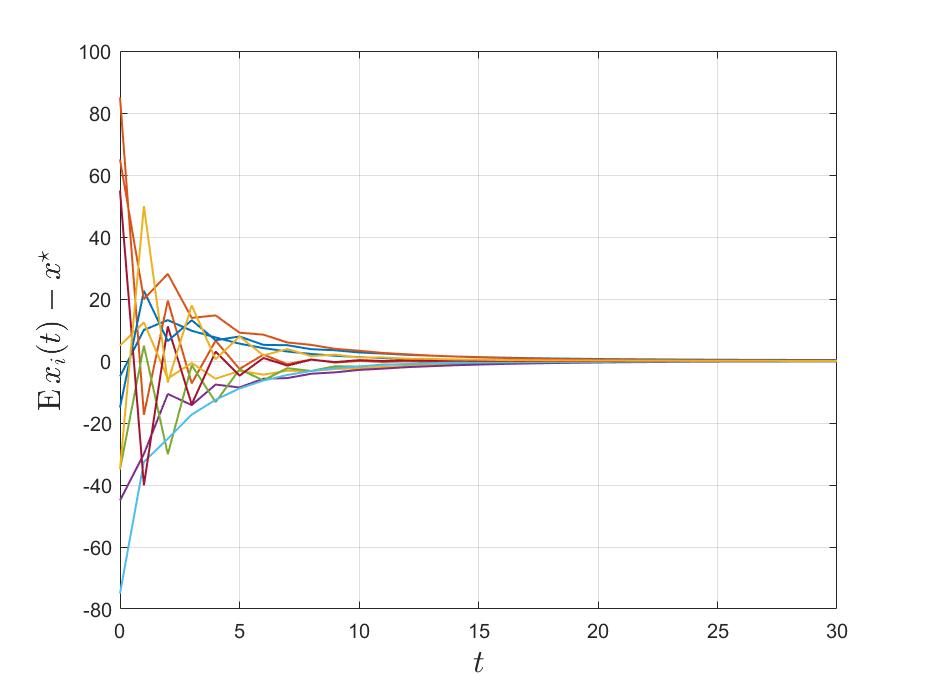}
	\caption{Trajectories of $\mathbb{E}[x_i(t)-x^{\star}]$ with Gaussian noises (500 samples).}
	\label{fig:G_E_x_1}
\end{figure}

\medskip 
\noindent{\bf Experiment 1.} (Gaussian Mechanism) First of all, we investigate the Gaussian mechanism with the expected differential privacy budget $(\epsilon,\delta)=(10^0,10^{-2})$ under a fixed adjacency level $\mu=1$. Let the message perturbation noise $\gamma_i(t) = \sigma_i\etab_i$ with $\etab_i\sim\mathcal{N}(0,1)$ and $\sigma_i=2.5244$, $i=1,\ldots,10$ by (\ref{eq:dp-gau-ac}). We  implement the algorithm for 500 times with the mean of the resulting computation errors given in Fig. \ref{fig:G_E_x_1}, which validates the conclusion that $\lim\limits_{t\rightarrow\infty}\mathbb{E}\big[\xb(t) -  \mathbf{1}_n\mathbf{1}_n^\top\xb(0)/n \big]=0$ in Proposition \ref{Theo-ConAcu}.(ii)  under  Gaussian noises.

%\begin{figure}[ht]
%	\hspace*{0cm}
%	\vspace*{0cm}
%	\centering
%	\includegraphics[width=8.6cm]{fig/fig_G_x_1.jpg}
%	\caption{Node state trajectories $x_i(t)$ with Gaussian noises.}
%	\label{fig:G_x_1}
%\end{figure}

\begin{figure}[ht]
	\hspace*{0cm}
	\vspace*{0cm}
	\centering
	\includegraphics[width=8cm]{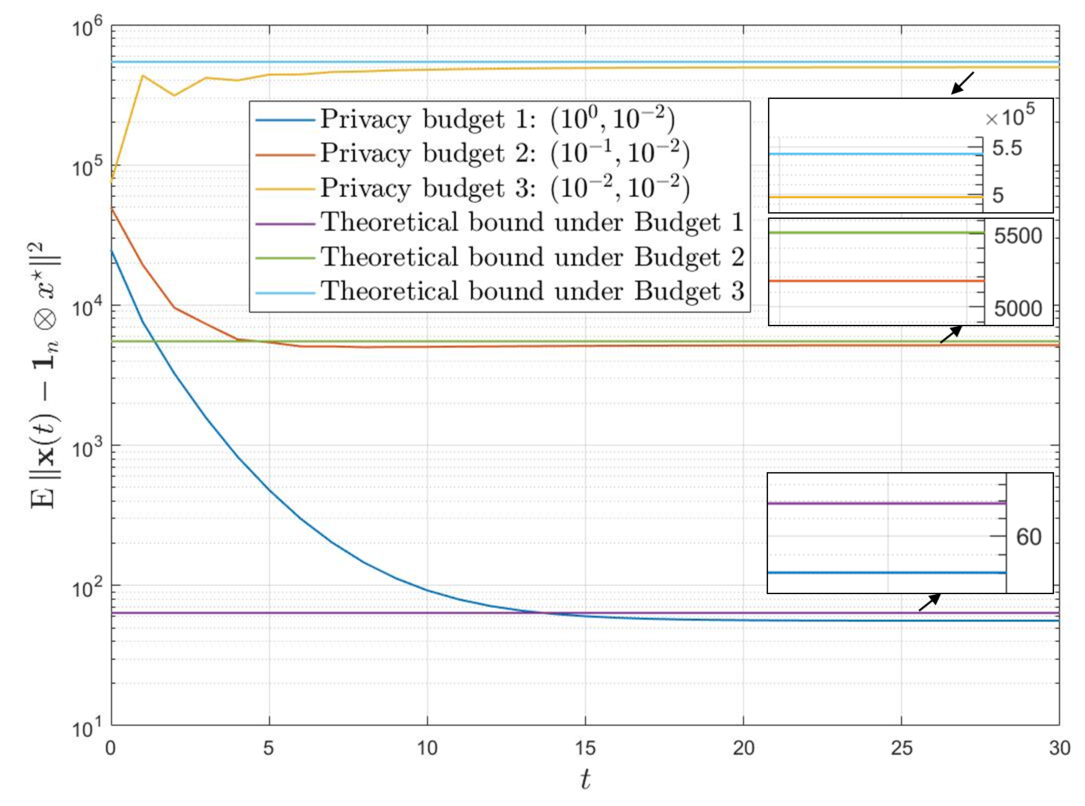}
	\caption{Trajectories of $\mathbb{E}\|\xb(t)-\mathbf{1}_n\otimes x^{\star}\|^2$ with Gaussian noises (500 samples).}
	\label{fig:G_MSE_total}
\end{figure}

Furthermore, we implement the algorithm for 500 times under three pairs of privacy budgets $(\epsilon,\delta)=(10^0,10^{-2}), (10^{-1},10^{-2}), (10^{-2},10^{-2})$ which correspond to $\sigma_i=2.5244, 23.4765, 232.8495$, respectively, and the resulting mean-square error trajectories are given in Fig. \ref{fig:G_MSE_total}. It can be seen that under all these three cases, the obtained mean-square errors in steady state are lower than the theoretical upper bounds by Proposition \ref{Theo-ConAcu}.(iii).

\begin{figure}[ht]
	\hspace*{0cm}
	\vspace*{0cm}
	\centering
	\includegraphics[width=8cm]{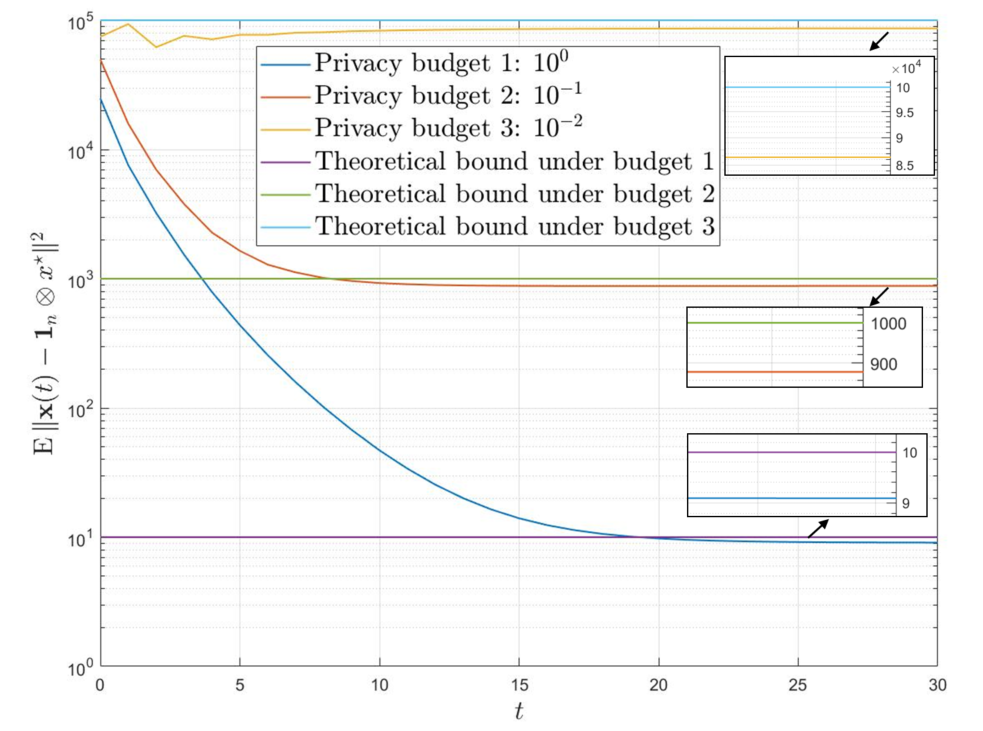}
	\caption{Trajectories of $\mathbb{E} \|\xb(t)-\mathbf{1}_n\otimes x^{\star}\|^2$ with Laplace noises (500 samples).}
	\label{fig:L_MSE_total}
\end{figure}

\medskip 
\noindent{\bf Experiment 2.} (Laplace Mechanism) Next, we move on to the Laplace mechanism with the expected differential privacy budget $\epsilon=10^0$ under a fixed adjacency level $\mu=1$. Let the message perturbation noise $\gamma_i(t) = \sigma_i\etab_i$ with $\etab_i\sim\mathcal{L}(0,1)$  for $i=1,\ldots,10$.  We   implement the algorithm for 500 independent rounds under three  privacy budgets $\epsilon=10^0, 10^{-1}, 10^{-2}$ which correspond to $\sigma_i=10^0, 10^{1}, 10^{2}$, respectively.
In Fig. \ref{fig:L_MSE_total}, it is shown that all resulting computation errors in the mean square sense are  smaller than the theoretical upper bounds, which, together with the above simulation results, validates Proposition \ref{Theo-ConAcu} under  Laplace noises.
Clearly from Fig. \ref{fig:G_MSE_total} and \ref{fig:L_MSE_total},  a higher privacy requirement (i.e., a smaller $\epsilon$) leads to a larger computation error in the mean-square sense, demonstrating the trade-off between the privacy level and the computation accuracy.

%\begin{figure}[ht]
%	\hspace*{0cm}
%	\vspace*{0cm}
%	\centering
%	\includegraphics[width=7cm]{fig/fig_L_x_1}
%	\caption{Node state trajectories of $x_i(t)- x^{\star}$ with Laplace noises.}
%	\label{fig:L_x_1}
%\end{figure}

%\begin{figure}[ht]
%	\hspace*{0cm}
%	\vspace*{0cm}
%	\centering
%	\includegraphics[width=7cm]{fig/fig_L_E_x_1}
%	\caption{Trajectories of $\mathbb{E} [x_i(t)- x^{\star}]$ with Laplace noises (500 samples).}
%	\label{fig:L_E_x_1}
%\end{figure}

\section{Conclusions}\label{secconc}
We have studied differential privacy for mechanisms generated by linear queries over  affine manifolds. We established necessary and sufficient conditions on whether differential privacy can be achieved when the affine manifolds encode geometry of the entries in a dataset.  
Finally, the derived framework was shown to be able to be applied to problems in  distributed consensus seeking and differentially private control, for which the affine-manifold constraint has been encoded in the systems themselves.  In future work, it would be interesting to further investigate extending these results to general  differentially private  mechanisms with nonlinear queries, and the resulting theories will then have direct  applicability in applications  such as differentially private deep learning. Besides, it is known that  the sequential composition theorem of $(\epsilon,\delta)$-differential privacy for multiple queries does not lead to a tighter quantitative result, and other notions such as R\'enyi differential privacy  and zero-concentrated differential privacy   have been introduced to improve such limitation. In view of this, another interesting future research direction is to generalize the established results to  R\'enyi differential privacy  and zero-concentrated differential privacy   for a tighter analysis of multiple queries over affine manifold.

\appendix
 
\section*{Appendix}
\section{Proof of Theorem \ref{Theo:BDP-Gau}}

\subsection{A Technical Lemma}
In this subsection, we present the necessary and sufficient condition from the perspective of privacy loss \cite{balle2018improving} for $(\epsilon,\delta)$-differential privacy of the mechanism \eqref{eq:mech-M} over the affine manifold $\mathcal{C}_d$.

Given any $\xb,\xb'$, we denote the random variables $\Yb = \mathscr{M}(\xb)$ and $\Yb' = \mathscr{M}(\xb')$, whose probability density function are given by $g(\yb-\Fb\xb)$ and $g(\yb'-\Fb\xb')$, respectively. Then, we define the privacy loss random variables $L_{\xb,\xb'} = \ell_{\xb,\xb'}(\Yb)$ and $L_{\xb',\xb} = \ell_{\xb',\xb}(\Yb')$,
where $\ell_{\xb,\xb'}$ is the privacy loss function of mechanism $\mathscr{M}$ on a pair of adjacent $\xb,\xb'$ as
\[
\ell_{\xb,\xb'}(\yb) = \log\left(\frac{g(\yb-\Fb\xb)}{g(\yb-\Fb\xb')}\right)\,.
\]

In view of the previous definitions, the mechanism output random variable $\Yb= \mathscr{M}(\xb)$ is transformed into the privacy loss random variable $L_{\xb,\xb'}$. Moreover, following \cite[Theorem 5]{balle2018improving}, the $(\epsilon,\delta)$-differential privacy over $\mathcal{C}_d$ can be rewritten in the form of the privacy loss random variable.
\begin{lemma}\label{lemma-app}
  The mechanism $\mathscr{M}$ is $(\epsilon,\delta)$-DP under $\mu$-adjacency over the affine manifold $\mathcal{C}_d$ if and only if
  \begin{equation}
  \label{eq:PLF-DP}
  \mathbb{P}\big(L_{\xb,\xb'}\geq \epsilon\big) - e^{\epsilon}\mathbb{P}\big(L_{\xb',\xb}\leq -\epsilon\big) \leq \delta
  \end{equation}
  holds for every pair $(\xb,\xb')\in\textnormal{Adj}(\mu,\mathcal{C}_{d})$.
\end{lemma}

\subsection{Proof of Necessity}
\label{sec-app-eq8}

In the following, we suppose that the mechanism $\mathscr{M}$ is $(\epsilon,\delta)$-differentially private  with privacy levels $\epsilon,\delta,\mu$, and  prove that both (\ref{eq:qual-bdp-G}) and \eqref{eq:quan-bdp-G} hold, respectively.

\noindent{\emph{Necessity of (\ref{eq:qual-bdp-G})}.}
We first apply the contradiction method to show that (\ref{eq:qual-bdp-G}) holds.
We suppose that (\ref{eq:qual-bdp-G}) is not satisfied, i.e.,
\begin{equation}\label{eq:theo-2-contra}
\rank(\Lambda) \neq \rank\left(\begin{bmatrix}\Lambda & \Fb\Db^{\bot}\end{bmatrix}\right)\,.
\end{equation}
Then we let $\Lambda_{\ker}\in\mathbb{R}^{(m-r)\times m}$ be such that $\Lambda_{\ker}\Lambda=0$ and $\rank\big([\Lambda^\dag;\, \Lambda_{\ker}]\big) = m$. Thus, we have $\Lambda_{\ker}\Fb\Db^{\bot}\neq0$, i.e., there exist $(\xb,\xb')\in\textnormal{Adj}(\mu,\mathcal{C}_{d})$ such that
\[
\Db(\xb-\xb')=0\,,\qquad \Lambda_{\ker}\Fb(\xb-\xb')\neq0\,.
\]

With the pair $(\xb,\xb')$ being the case, we let $\mathcal{M}_\ast\subset\mathbb{R}^m$ such that $\Lambda_{\ker}\Fb\xb\in \Lambda_{\ker} \mathcal{M}_\ast$, $\Lambda_{\ker}\Fb\xb'\notin \Lambda_{\ker} \mathcal{M}_\ast$ and $\Lambda^\dag \mathcal{M}_\ast = \mathbb{R}^{r}$. Then it is observed that
\[
\ba{rcl}
&&\mathbb{P}\big(\mathscr{M}(\xb) \in \mathcal{M}_\ast \big) \\
&=& \mathbb{P}\big(\Lambda_{\ker}\Fb\xb \in \Lambda_{\ker} \mathcal{M}_\ast, \Lambda^\dag \Fb\xb + \Lambda^\dag\Lambda\etab\in \Lambda^\dag \mathcal{M}_\ast\big)\\
&=& \mathbb{P}\big(\Lambda_{\ker}\Fb\xb \in \Lambda_{\ker} \mathcal{M}_\ast\big) \mathbb{P}\big(\Lambda^\dag \Fb\xb + \etab\in \mathbb{R}^{r}\big)\\
&=& 1
\ea
\]
and similarly,
\[
\ba{rcl}
&&\mathbb{P}\big(\mathscr{M}(\xb') \in \mathcal{M}_\ast \big)\\
&=& \mathbb{P}\big(\Lambda_{\ker}\Fb\xb' \in \Lambda_{\ker} \mathcal{M}_\ast\big) \mathbb{P}\big(\Lambda^\dag \Fb\xb' + \etab\in \mathbb{R}^{r}\big)\\
&=& 0
\ea
\]
This indicates that there is no $(\epsilon,\delta)$ such that \eqref{eq:DP} is satisfied, contradicting with the fact that $\mathscr{M}$ is differentially privat  with privacy levels $(\epsilon,\delta,\mu)$. Therefore, \eqref{eq:theo-2-contra} does not hold, which completes the necessity proof of statement (i).

\medskip
\noindent{\emph{Necessity of (\ref{eq:quan-bdp-G})}.}
In view of the previous analysis, for any $\xb\in\mathbb{R}^n$ and any $\mathcal{M}\subseteq\mathbb{R}^m$ it follows that
\begin{equation}
    \label{eq:PM}
    \ba{rcl}
  \mathbb{P}\big(\mathscr{M}(\xb) \in \mathcal{M} \big)&=&  \mathbb{P}\big(\Lambda_{\ker}\Fb\xb \in \Lambda_{\ker} \mathcal{M}\big)\\ &&\times \mathbb{P}\big(\Lambda^\dag \Fb\xb + \etab\in \Lambda^\dag \mathcal{M}\big)\,.
  \ea
\end{equation}
With this in mind, we suppose that $\mathscr{M}$ over $\mathcal{C}_d$ is $(\epsilon,\delta)$-differentially private and (\ref{eq:qual-bdp-G}) is satisfied, and proceed to show the inequality (\ref{eq:quan-bdp-G}) is also satisfied.

We first show
\begin{equation}\label{eq:PM1}
\mathbb{P}\big(\Lambda_{\ker}\Fb\xb \in  \mathcal{M}_1\big) = \mathbb{P}\big(\Lambda_{\ker}\Fb\xb' \in \mathcal{M}_1\big) \in \{0,1\}
\end{equation}
for all $\mathcal{M}_1\subseteq\mathbb{R}^{m-r}$ and all ${(\xb,\xb')\in{\rm Adj}(\mu,\mathcal{C}_d)}$.
By the definition of $\mathcal{C}_d$, given any ${(\xb,\xb')\in{\rm Adj}(\mu,\mathcal{C}_d)}$ it is clear that $\Db (\xb-\xb') = 0$. On the other hand, by (\ref{eq:qual-bdp-G}) we have $\Lambda_{\ker}\Fb\Db^{\bot}=0$. Thus, it can be easily verified that $\Lambda_{\ker}\Fb(\xb-\xb') = 0$ for all ${(\xb,\xb')\in{\rm Adj}(\mu,\mathcal{C}_d)}$. This immediately yields (\ref{eq:PM1}) by recalling that the event $\Lambda_{\ker}\Fb\xb \in  \mathcal{M}_1$ is deterministic.

With (\ref{eq:PM1}), we now proceed to show (\ref{eq:quan-bdp-G}). By combining (\ref{eq:PM}) and (\ref{eq:PM1}), given any ${(\xb,\xb')\in{\rm Adj}(\mu,\mathcal{C}_d)}$ we have
\begin{equation}\label{eq:DP-equi}\ba{rcl}
 &\mathbb{P}\big(\mathscr{M}(\xb) \in \mathcal{M} \big) \leq e^{\epsilon}\mathbb{P}\big(\mathscr{M}(\xb') \in \mathcal{M} \big)+\delta\,,\\ & \qquad \forall \mathcal{M}\subseteq\mathbb{R}^m\\
 \iff& \mathbb{P}\big(\Lambda^\dag \Fb\xb + \etab \in \mathcal{M}_2 \big) \leq e^{\epsilon}\mathbb{P}\big(\Lambda^\dag \Fb\xb' + \etab \in \mathcal{M}_2 \big)+\delta\,,\\ & \qquad \forall \mathcal{M}_2\subseteq\mathbb{R}^r
\ea
\end{equation}
It is noted that the lower inequality of (\ref{eq:DP-equi}) indicates that the mechanism  $\mathscr{M}_2(\xb):=\Lambda^\dag \Fb\xb + \etab$ over $\mathcal{C}_d$ is $(\epsilon,\delta)$-differentially private.

Next, we apply this fact and Lemma \ref{lemma-app} to the mechanism  $\mathscr{M}_2(\xb)$ to show (\ref{eq:quan-bdp-G}).
It can be verified that the random variables $\Yb:=\Lambda^\dag \Fb\xb + \etab\backsim\mathcal{N}(\Lambda^\dag \Fb\xb,\Ib_r)$, $\Yb':=\Lambda^\dag \Fb\xb' + \etab\backsim\mathcal{N}(\Lambda^\dag \Fb\xb',\Ib_r)$ and the privacy loss function
\[
\ell_{\xb,\xb'}(\yb)= (\yb-\Lambda^\dag \Fb\xb)^\top\Lambda^\dag \Fb(\xb-\xb') + \frac{1}{2}\|\Lambda^\dag \Fb(\xb-\xb')\|^2
\]
Then, we can obtain both the privacy loss random variables $L_{\xb,\xb'}$ and $L_{\xb',\xb}$ share the same distributions as
\[
L_{\xb,\xb'}\backsim\mathcal{N}(\eta/2,\eta)
\]
with $\eta=\|\Lambda^\dag \Fb(\xb-\xb')\|^2$. According to Lemma \ref{lemma-app}, the  $(\epsilon,\delta)$-differential privacy of $\mathscr{M}_2$ is equivalent to saying
\[
\mathbb{P}\big(L_{\xb,\xb'}\geq \epsilon\big) - e^{\epsilon}\mathbb{P}\big(L_{\xb',\xb}\leq -\epsilon\big) \leq \delta
\]
i.e.,
\begin{equation}\label{eq:Phi-eta}
\Phi(-\frac{\epsilon}{\sqrt{\eta}} + \frac{\sqrt{\eta}}{2}) - e^{\epsilon}\Phi(-\frac{\epsilon}{\sqrt{\eta}} - \frac{\sqrt{\eta}}{2}) \leq \delta
\end{equation}
for all ${(\xb,\xb')\in{\rm Adj}(\mu,\mathcal{C}_d)}$. We note that the left side of the above inequality increases as $\eta$ increases by \cite{balle2018improving}. Thus, by defining the $L_2$ sensitivity of mechanism $\mathscr{M}_2$ as
\[
\Delta_2 := \sup_{(\xb,\xb')\in{\rm Adj}(\mu,\mathcal{C}_d)}\|\Lambda^\dag \Fb(\xb-\xb')\|
\]
there must hold
\begin{equation}\label{eq:etamax}
\Phi(-\frac{\epsilon}{\Delta_2} + \frac{\Delta_2}{2}) - e^{\epsilon}\Phi(-\frac{\epsilon}{\Delta_2} - \frac{\Delta_2}{2}) \leq \delta\,.
\end{equation}

We now proceed to show that $\Delta_2= \mu\Delta_{\mathcal{N}}$. For any ${(\xb,\xb')\in{\rm Adj}(\mu,\mathcal{C}_d)}$, by recalling \eqref{eq:xxprime} and the definition of $\Psi_j$ at the beginning of Section \ref{sec-IIIA}, it follows
\[
\Delta_2=  \max_{i\in{\rm V},j\in\mathcal{I}}\|\Lambda^\dag\Fb\Psi_j\eb_i\|\mu = \mu \max_{i\in{\rm V}}\Delta_i^{\mathcal{N}} = \mu\Delta_{\mathcal{N}}\,.
\]
This thus proves  (\ref{eq:quan-bdp-G}) by (\ref{eq:etamax}) and \eqref{eq:kappa}.

\subsection{Proof of Sufficiency}

With \eqref{eq:qual-bdp-G} and \eqref{eq:quan-bdp-G}, we now prove $(\epsilon,\delta)$-differential privacy of mechanism $\mathscr{M}$   over $\mathcal{C}_d$.

We first show that under \eqref{eq:quan-bdp-G} the mechanism $\mathscr{M}_2(\xb):=\Lambda^\dag \Fb\xb + \etab$ over $\mathcal{C}_d$ is $(\epsilon,\delta)$-differentially private.
According to Lemma \ref{lemma-app}, this is equivalent to proving \eqref{eq:Phi-eta} for all ${(\xb,\xb')\in{\rm Adj}(\mu,\mathcal{C}_d)}$. Then, by recalling {that the $L_2$ sensitivity of mechanism $\mathscr{M}_2$ satisfies $\Delta_2= \mu\Delta_{\mathcal{N}}$}, one can immediately conclude from \eqref{eq:quan-bdp-G} that \eqref{eq:Phi-eta} is true for all ${(\xb,\xb')\in{\rm Adj}(\mu,\mathcal{C}_d)}$, i.e.,
\[
\mathbb{P}\big(\Lambda^\dag \Fb\xb + \etab \in \mathcal{M}_2 \big) \leq e^{\epsilon}\mathbb{P}\big(\Lambda^\dag \Fb\xb' + \etab \in \mathcal{M}_2 \big)+\delta
\]
for all $\mathcal{M}_2\subseteq\mathbb{R}^r$. 
Thus, by (\ref{eq:PM}) and (\ref{eq:PM1}) we further have
\[\ba{rcl}
&&\mathbb{P}\big(\mathscr{M}(\xb) \in \mathcal{M} \big) \\
&=&  \mathbb{P}\big(\Lambda_{\ker}\Fb\xb \in \Lambda_{\ker} \mathcal{M}\big) \mathbb{P}\big(\Lambda^\dag \Fb\xb + \etab\in \Lambda^\dag \mathcal{M}\big)\\
&\leq & \mathbb{P}\big(\Lambda_{\ker}\Fb\xb' \in \Lambda_{\ker}\mathcal{M}\big) \Big(e^{\epsilon}\mathbb{P}\big(\Lambda^\dag \Fb\xb' + \etab \in \Lambda^\dag \mathcal{M} \big)+\delta\Big)\\
&\leq& e^{\epsilon}\mathbb{P}\big(\Lambda_{\ker}\Fb\xb' \in \Lambda_{\ker}\mathcal{M}\big)\mathbb{P}\big(\Lambda^\dag \Fb\xb' + \etab \in \Lambda^\dag \mathcal{M} \big) +\delta\\
&=& e^{\epsilon}\mathbb{P}\big(\mathscr{M}(\xb') \in \mathcal{M} \big) +\delta\,
\ea\]
for all $\mathcal{M}\subseteq\mathbb{R}^m$.
Therefore, the proof is completed.

\section{Proof of Theorem \ref{Theo:BDP-Lap}}

\subsection{Proof of Necessity}

{\emph{Necessity of (\ref{eq:qual-bdp-G})}.} The proof is the same as the necessity proof of (\ref{eq:qual-bdp-G}), and is thus omitted for simplicity.

\medskip
\noindent
{\emph{Necessity of (\ref{eq:quan-bdp-Lap})}.}
Similar to the arguments in the necessity proof of (\ref{eq:quan-bdp-G}), we can obtain (\ref{eq:DP-equi}) and thus the $(\epsilon,0)$-differential privacy of the Laplace mechanism $\mathscr{M}_2(\xb):=\Lambda^\dag \Fb\xb + \etab$ w.r.t. $\mathcal{C}_d$ for $\etab\sim\mathcal{L}(0,1)^r$, by using (\ref{eq:qual-bdp-G}) and the fact that $\mathscr{M}(\xb)$ is $(\epsilon,0)$-differentially private over $\mathcal{C}_d$.  With this in mind, we then proceed to  show (\ref{eq:quan-bdp-Lap}) by contradiction.

We suppose $\Delta_{\mathcal{L}} > \epsilon/\mu$. This, by the definition of $\Delta_{\mathcal{L}}$ and (\ref{eq:xxprime}), indicates that there exists ${(\xb,\xb')\in{\rm Adj}(\mu,\mathcal{C}_d)}$ such that
\[
\|\Lambda^\dag\Fb(\xb-\xb')\|_1 = \mu \Delta_{\mathcal{L}} > \epsilon\,.
\]
With such $(\xb,\xb')$ being the case, we then can always construct a non-empty set $\mathcal{S}\subset\mathbb{R}^r$ such that
\[\ba{rcl}
\|\ub - \Lambda^\dag\Fb\xb\|_1 &=& \|\ub - \Lambda^\dag\Fb\xb'-\Lambda^\dag\Fb(\xb-\xb')\|_1\\
&=& \|\ub - \Lambda^\dag\Fb\xb'\|_1 -\|\Lambda^\dag\Fb(\xb-\xb')\|_1
\ea\]
for all $\ub\in\mathcal{S}$.
We then let $\mathcal{M}_2=\{\yb\in\mathbb{R}^r: \yb=\ub-\Lambda^\dag\Fb\xb, \ub\in\mathcal{S}\}$, and observe that
\[
\ba{rcl}
&&\mathbb{P}\big(\Lambda^\dag \Fb\xb + \etab \in \mathcal{M}_2 \big) \\
&=& \frac{1}{2^r}\int_{\mathcal{M}_2}e^{-\|\etab\|_1}d\etab\\
&=& \frac{1}{2^r}\int_{\mathcal{S}}e^{-\|\ub-\Lambda^\dag\Fb\xb\|_1}d\ub\\
&=& \frac{1}{2^r}\int_{\mathcal{S}}e^{-\|\ub-\Lambda^\dag\Fb\xb'\|_1+\|\Lambda^\dag\Fb(\xb-\xb')\|_1}d\ub\\
&=& e^{\|\Lambda^\dag\Fb(\xb-\xb')\|_1} \mathbb{P}\big(\Lambda^\dag \Fb\xb' + \etab \in \mathcal{M}_2 \big)\\
&>& e^{\epsilon} \mathbb{P}\big(\Lambda^\dag \Fb\xb' + \etab \in \mathcal{M}_2 \big),
\ea
\]
which clearly contradicts with the fact that $\mathscr{M}_2$ is $(\epsilon,0)$-differentially private, i.e., for all $\mathcal{M}_2\subseteq\mathbb{R}^r$ and all ${(\xb,\xb')\in{\rm Adj}(\mu,\mathcal{C}_d)}$, there holds
\[
\mathbb{P}\big(\Lambda^\dag \Fb\xb + \etab \in \mathcal{M}_2 \big) \leq e^{\epsilon}\mathbb{P}\big(\Lambda^\dag \Fb\xb' + \etab \in \mathcal{M}_2 \big)\,.
\]
Therefore, we have $\Delta_{\mathcal{L}} \leq \epsilon/\mu$, proving (\ref{eq:quan-bdp-Lap}).

\subsection{Proof of Sufficiency}

With \eqref{eq:qual-bdp-G} and (\ref{eq:quan-bdp-Lap}), we now prove $(\epsilon,0)$-differential privacy of  mechanism $\mathscr{M}$   over $\mathcal{C}_d$.

We first show that under (\ref{eq:quan-bdp-Lap}) the Laplace mechanism $\mathscr{M}_2(\xb):=\Lambda^\dag \Fb\xb + \etab$ over $\mathcal{C}_d$ is $(\epsilon,0)$-differentially private. We first observe from the definition of $\Delta_{\mathcal{L}}$ and (\ref{eq:xxprime}) that the $L_1$ sensitivity of $\mathscr{M}_2$ satisfies
\[\ba{rcl}
\Delta_1&:=&\sup_{(\xb,\xb')\in{\rm Adj}(\mu,\mathcal{C}_d)}\|\Lambda^\dag\Fb(\xb-\xb')\|_1\\
&=&\max_{i\in{\rm V},j\in\mathcal{I}}\|\Lambda^\dag\Fb\Psi_j\eb_i\|_1\mu \\
&=& \mu \Delta_{\mathcal{L}} \leq \epsilon\,.
\ea\]
Then, we note that for all $\mathcal{M}_2\subseteq\mathbb{R}^r$,
\[\ba{rcl}
\mathbb{P}\big(\Lambda^\dag \Fb\xb + \etab \in \mathcal{M}_2 \big) &\leq& e^{\|\Lambda^\dag\Fb(\xb-\xb')\|_1}\mathbb{P}\big(\Lambda^\dag \Fb\xb + \etab \in \mathcal{M}_2 \big)\\
&\leq& e^{\epsilon}\mathbb{P}\big(\Lambda^\dag \Fb\xb + \etab \in \mathcal{M}_2 \big)\,,
\ea\]
proving $(\epsilon,0)$-differential privacy of $\mathscr{M}_2$ over the affine manifold $\mathcal{C}_d$.
Thus, by (\ref{eq:PM}) and (\ref{eq:PM1}) we further have
\[\ba{rcl}
&&\mathbb{P}\big(\mathscr{M}(\xb) \in \mathcal{M} \big)\\
&=&  \mathbb{P}\big(\Lambda_{\ker}\Fb\xb \in \Lambda_{\ker} \mathcal{M}\big) \mathbb{P}\big(\Lambda^\dag \Fb\xb + \etab\in \Lambda^\dag \mathcal{M}\big)\\
&\leq & \mathbb{P}\big(\Lambda_{\ker}\Fb\xb' \in \Lambda_{\ker}\mathcal{M}\big) \Big(e^{\epsilon}\mathbb{P}\big(\Lambda^\dag \Fb\xb' + \etab \in \Lambda^\dag \mathcal{M} \big)\Big)\\
&=& e^{\epsilon}\mathbb{P}\big(\mathscr{M}(\xb') \in \mathcal{M} \big)\,.
\ea\]
for all $\mathcal{M}\subseteq\mathbb{R}^m$.
Therefore, the proof is completed.

\section{Proof of Theorem 3}
\label{app:proof_lemma1}

To prove Theorem 3, we first need to further elaborate the expressions of  $N_{ij}$ in (\ref{eq:bar-N_ij})  with the mechanism $\mathscr{M}$ in (\ref{eq:mechanism-cont}) and the  manifold $\mathcal{C}_d$ in (\ref{eq:manifold-cont}). It is observed that with $\Db$ in (\ref{eq:Db}), there exist $T$ sets $\mathrm{d}_j$, $j\in\mathcal{I}:=\{1,\ldots,T\}$ such that $\Db_{\mathrm{d}_j}\in\mathbb{R}^{(n-n_x)\times(n-n_x)}$ is nonsingular and $-\mathrm{d}_j:=\mathrm{V}/\mathrm{d}_j=\{(j-1)n_x+1,\ldots,(j-1)n_x+n_x\}$. It then  follows that
\[
\ba{rcl}
\Psi_j\Eb_{-\mathrm{d}_j} &=& \big(\Ib_n - \mathbf{E}_{\mathrm{d}_j}(\Db_{\mathrm{d}_j})^{-1}\Db\big)\Eb_{-\mathrm{d}_j}\\
&=& \Eb_{-\mathrm{d}_j} - \mathbf{E}_{\mathrm{d}_j}(\Db_{\mathrm{d}_j})^{-1}\Db_{-\mathrm{d}_j}\\
&=& \Db^\bot\Ab^{1-j}
\ea
\]
where the last equality is obtained by using $(\Db_{\mathrm{d}_j})^{-1}\Db_{-\mathrm{d}_j}=\big[\Ab^{1-j};\ldots;\Ab^{-1};\Ab;\ldots;\Ab^{T-j}\big]$ and the definition of $\Db^\bot$ in (\ref{eq:D-bot}). This further implies $\bar\Lambda^{\dag}\Fb\Psi_j\Eb_{-\rm d_j}\Eb_{-\rm d_j}^\top\eb_i=\bar\Lambda^{\dag}\mathbf{O}_T\Ab^{1-j}\Eb_{-\rm d_j}^\top\eb_i$ for $i\in\mathrm{V}$ and $j\in\mathcal{I}$.

Moreover, we have
\[\ba{rcl}
N_{ij}&:=& \bar\Lambda^{\dag}\Fb\Psi_j\Eb_{-\rm d_j}\Eb_{-\rm d_j}^\top\eb_i \\
&=&\bar\Lambda^{\dag}\mathbf{O}_T\Ab^{1-j}\Eb_{-\rm d_j}^\top\eb_i\,\\
&=&\Ab^{1-j}\Eb_{-\rm d_j}^\top\eb_i
\ea\]
for all $(i,j)\in\mathrm{V}\times\mathcal{I}$.
It is also noted that for any $i\in-\mathrm{d}_j$, $\Eb_{-\mathrm{d}_j}^\top \mathbf{e}_i=\mathbf{v}_k$ with $k:=i-(j-1)n_x\in[1,n_x]$, while for any $i\in\mathrm{d}_j$, $\Eb_{-\mathrm{d}_j}^\top \mathbf{e}_i=0$.
It thus follows that \eqref{eq:dp-gau-ac-2} implies \eqref{eq:quan-bdp-GG}, completing the proof.

\section{Proof of Proposition \ref{Theo-ConAcu}}

According to the algorithm (\ref{eq:AveCon}) and the noise design in Algorithm 1, we first introduce the evolution equation of the network node state $\xb(t)$, as
\begin{equation}\label{eq:xb}
  \xb(t+1) = (\Ib_n-\Lb)\xb(t) - \Lb\Delta_{\sigma}\etab
\end{equation}
where  $\Delta_{\sigma}=\mbox{diag}(\sigma_1,\ldots,\sigma_n)$ and $\etab=[\etab_1;\ldots;\etab_n]$. By multiplying both sides of (\ref{eq:xb}) with $\mathbf{1}_n^\top$, one can obtain
 \[
 \mathbf{1}_n^\top\xb(t+1) = \mathbf{1}_n^\top(\Ib_n-\Lb)\xb(t) - \mathbf{1}_n^\top\Lb\Delta_{\sigma}\etab = \mathbf{1}_n^\top\xb(t)
 \]
 where the last equality is derived by using $\mathbf{1}_n^\top\Lb=0$. This thus implies that the average of the node states is preserved, i.e., $\mathbf{1}_n^\top\xb(t)=\mathbf{1}_n^\top\xb(0)$.
 Then denote the error between the node states $x_i$ and the average of initial states as $z_i(t) = x_i(t) - \mathbf{1}_n^\top\xb(0)/n$, and let $\zb(t)=[z_1(t);\ldots;z_n(t)]$, whose evolution satisfies
\[\ba{rcl}
  \zb(t+1) &=& (\Ib_n-\Lb-\frac{1}{n}\mathbf{1}_n\mathbf{1}_n^\top)\xb(t) - \Lb\Delta_{\sigma}\etab \\
  &=& (\Ib_n-\Lb-\frac{\lambda_2}{n}\mathbf{1}_n\mathbf{1}_n^\top)\zb(t) - \Lb\Delta_{\sigma}\etab\,
\ea\]
where to derive the second equality we have used the equality $(\Ib_n-\Lb-\frac{\lambda_2}{n}\mathbf{1}_n\mathbf{1}_n^\top)(\Ib_n-\frac{1}{n}\mathbf{1}_n\mathbf{1}_n^\top)=(\Ib_n-\Lb-\frac{1}{n}\mathbf{1}_n\mathbf{1}_n^\top)$.
Thus, we have
\begin{equation}\label{eq:zb}
\zb(t) = (\Ib_n-\Lb-\frac{\lambda_2}{n}\mathbf{1}_n\mathbf{1}_n^\top)^t\zb(0) - [\Ib_n-(\Ib_n-\Lb)^t]\Delta_{\sigma}\etab\,.
\end{equation}
Note that $\|\Ib_n-\Lb-\frac{\lambda_2}{n}\mathbf{1}_n\mathbf{1}_n^\top\| \leq \alpha$ and $\|\Ib_n-(\Ib_n-\Lb)^t\|\leq 1+\alpha^t$ with $\alpha:=\max\{|1-\lambda_2|,|1-\lambda_n|\}\in(0,1)$.
Now we proceed to analyze the convergence property of $\zb(t)$, and observe that 
\[\ba{rcl}
\|\zb(t)\| 
\leq  \alpha^t\|\xb(0)\| + (1+\alpha^t)\big(\sum_{i\in\mathrm{V}}\sigma_i^2\etab_i^2\big)^{1/2}
\ea\]
which proves the first statement by recalling $\zb(t):=\xb(t) -  \mathbf{1}_n\mathbf{1}_n^\top\xb(0)/n$.

Then we analyze the computation accuracy by computing the mean and mean-square of $\zb(t)$ in (\ref{eq:zb}). As $\etab$ is a vector of mutually independent noises with zero mean and identity variance, it is clear that
\[\ba{rcl}
\mathbb{E}\zb(t) &=& (\Ib_n-\Lb-\frac{\lambda_2}{n}\mathbf{1}_n\mathbf{1}_n^\top)^t\zb(0) \\
\mathbb{E}\|\zb(t)\|^2 &=& \|(\Ib_n-\Lb-\frac{\lambda_2}{n}\mathbf{1}_n\mathbf{1}_n^\top)^t\zb(0)\|^2 \\&& + \mathbb{E}\|[\Ib_n-(\Ib_n-\Lb)^t]\Delta_{\sigma}\etab\|^2\\
&\leq& \alpha^{2t}\|\zb(0)\|^2  + (1+\alpha^{t})^2\sum_{i\in\mathrm{V}}\sigma_i^2\mathbb{E}\|\etab_i\|^2\,
\ea\]
with  $\|\Ib_n-\Lb-\frac{\lambda_2}{n}\mathbf{1}_n\mathbf{1}_n^\top\| \leq \alpha<1$. This thus completes the proof.

\bibliographystyle{agsm}
\bibliography{IEEE_TIFS}

% that's all folks
\end{document}